%% file: 3481.tex
\documentclass[twocolumn]{aa}
\usepackage{graphicx}
\usepackage{natbib}
\newcommand{\teff}{T$_{\rm eff}$}

\newcommand{\nodata}{\ldots}

\begin{document}
   \title{Mg isotopic ratios in giant stars of the globular cluster 
   NGC 6752\thanks{Based on observations obtained with 
   the ESO Very Large Telescope UVES spectrograph for programmes 67.D-0145 
   and 65.L-0165}}

\authorrunning{D. Yong et al.}
\titlerunning{Mg isotopic ratios in NGC 6752}

   \author{D. Yong
          \inst{1}
          \and
          F. Grundahl\inst{2}
          \and
          D. L. Lambert\inst{1}
          \and
          P. E. Nissen\inst{2}
          \and 
          M. D. Shetrone\inst{3}
          }

   \offprints{D. Yong}

   \institute{Department of Astronomy, University of Texas, Austin, TX 78712, USA \\
              \email{tofu,dll@astro.as.utexas.edu}
         \and
             Department of Physics and Astronomy, University of Aarhus, 8000 Aarhus C, Denmark\\
             \email{fgj,pen@phys.au.dk}
         \and
             McDonald Observatory, University of Texas, HC 75, Box 1337 MCD, Fort Davis, TX 79734, USA\\
             \email{shetrone@astro.as.utexas.edu}
             }

   \date{}

   \abstract{

Mg isotopic abundance ratios are measured in 20 bright red giants
in globular cluster NGC 6752 based on very high-resolution (R$\sim$110,000),
high signal-to-noise spectra obtained with UVES on the VLT.  
There is a considerable spread in the ratio $^{24}$Mg:$^{25}$Mg:$^{26}$Mg
with values ranging from 53:9:39 to 83:10:7.  
We measured the abundances of O, Na, Mg, Al, and Fe combining our sample
with 21 RGB bump stars \citep{grundahl02}.  The abundances of the
samples are consistent and exhibit the usual anticorrelations between O-Na and Mg-Al.  
A positive correlation is found between $^{26}$Mg and Al, a mild anticorrelation is
found between $^{24}$Mg and Al, while no correlation is found between $^{25}$Mg and Al.
None of the elemental or isotopic abundances show a 
dependence on evolutionary status and, as shown by \citet{gratton01}, the abundance
variations exist even in main sequence stars.  This strongly suggests that the 
star-to-star abundance variations are a result of varying degrees of pollution 
with intermediate mass AGB stars being likely polluters.  
Consideration of the extremes of the abundance variations
provides the composition of the ambient material and the processed material.
In the least contaminated stars (lowest Na and Al and highest O and Mg abundances), 
we infer a Mg isotopic ratio around 80:10:10 and a composition ([O/Fe], [Na/Fe], [Mg/Fe], [Al/Fe]) 
$\simeq$ (0.6 ,$-$0.1, 0.5, 0.0).
In the most polluted stars, we find a Mg isotopic ratio around 60:10:30 and a 
composition ([O/Fe], [Na/Fe], [Mg/Fe], [Al/Fe]) $\simeq$ ($-$0.1, 0.6, 0.3, 1.2).
Even for the least polluted stars, the abundances of $^{25}$Mg and 
$^{26}$Mg relative to $^{24}$Mg are considerably higher
than predicted for ejecta from $Z=0$ supernovae.  Zero metallicity AGB stars may be
responsible for these higher abundances.
Our measured Mg isotopic ratios reveal another layer to the globular cluster
star-to-star abundance variations that demands extensions of our present theoretical 
knowledge of stellar nucleosynthesis by giant stars.

   }

   \maketitle
%

\section{Introduction}

\label{sec:intro}

Galactic globular clusters continue to play a central role
in modern astrophysics for a variety of reasons.  Firstly, as the
oldest Galactic objects for which reliable ages have been determined \citep{vsb96},
they place a lower limit on the age of the universe.  Secondly, besides
the notable exceptions $\omega$ Cen and 
possibly M 22 \citep{lehnert91}, all stars within a given
cluster have the same iron abundance within a narrow range though the
metallicity varies considerably from cluster to cluster.  This is surely 
a vital clue regarding the origins of globular clusters that
has received relatively little theoretical attention 
\citep{brown91,brown95,lm96,parmentier99,nmn00}.
Finally, every well studied Galactic globular cluster shows star-to-star abundance
variations for light elements (C, N, O, Na, Mg, and
Al).  Although the amplitude of the variations may differ from
cluster to cluster, there is a common pattern: the abundances of C and O
are low when N is high, O and Na are anticorrelated as are Mg and Al.

Two scenarios compete for priority in accounting for the star-to-star 
abundance variations: the so-called evolutionary and primordial
scenarios.  The evolutionary scenario supposes that the variations
arise when a star becomes a red giant with a
convective envelope that may
tap deep layers where H-burning occurs through
the CNO-cycles, and possibly the Ne-Na
and Mg-Al chains.  Standard models of red giants do not predict the
required deep extension of the convective envelope but its
invocation is not totally physically implausible.  If the extension
were controlled by (say) the rotation of the star, a star-to star
abundance variation could occur among red giants.  A primordial
scenario places the first appearance of the abundance variations
in main sequence stars.  One possibility is that the cluster gas
was not of a homogeneous composition when the stars formed.  A
second possibility is that stars may have formed with identical or
nearly identical compositions but subsequent accretion of gas
ejected by evolved stars resulted in the star-to-star abundance
variations. 

Abundance variations were discovered first among cluster red giants
for obvious reasons.  Later, variations were found to be present
among subgiants and even in main sequence stars where an
evolutionary scenario is excluded on the firm ground that the
internal temperatures required for hydrostatic equilibrium are too
low for the Ne-Na and Mg-Al chains to run by proton captures.  The
pertinent observations involve variously: narrow band photometry and
low or high resolution spectroscopy 
including those reported initially by   
\citet{hesser78}, \citet{hesser80}, and \citet{bell83} for 47 Tuc
and more recently by \citet{cannon98}, \citet{cohen99}, \citet{gratton01}, 
\citet{cohen02}, \citet{briley02}, and \citet{ramirez03} for a variety
of clusters.
In some or all clusters, primordial variations for some elements may be supplemented
by an evolutionary component. For example,
the carbon abundances are correlated with luminosity of the
red giants of globular clusters M 4, M 22, NGC 6752, $\omega$ Cen,
and 47 Tuc \citep{smith89,brown89,ss91,cannon02}. 
It has also been suggested that variations in the He abundance 
may play a role in the morphology of the horizontal branch \citep{dantona02}.

In this paper, we examine a novel aspect of the star-to-star
abundance variations among red giants of the globular
cluster NGC 6752. 
This is the cluster in which \citet{cottrell81} discovered the
correlated Na and Al enhancements in CN-strong giants.  This cluster
with M 13 \citep{kraft97} shows the largest amplitude in the star-to-star
abundance variations of all the well-studied globular clusters.
\citet{ss91} through a
combination of low-resolution optical and infrared spectra found
C and N abundance variations extending from the main sequence to the
red giant stars of NGC 6752.  Among red giants, the carbon abundance was
shown to decline with increasing luminosity, a sign of an
evolutionary contribution to the predominant primordial
scenario for explaining the abundance variations. Recently,
\citet{gratton01} reported finding the O-Na and Mg-Al anticorrelations
among stars at the main sequence turn-off and \citet{grundahl02}
obtained similar results for stars on the red giant branch around the
bump on this branch. Our extension involves the measurement of
magnesium isotopic ratios for a sample of the most
luminous giants.  We also determine the O,  Na, Mg, Al, and Fe
abundances. Our goal is to examine the primordial and evolutionary
scenarios in light of the Mg isotopic abundances.  

The first attempt to measure Mg isotopic ratios for globular cluster
giants was by \citet{shetrone96a,shetrone96b} for a few giants in M13, 
a cluster with both a similar metallicity and spread in light element 
abundances to NGC 6752.  
With the VLT and UVES, it is possible to obtain the high quality
spectra necessary for measuring the Mg isotopic ratios in a sample of
the brightest giants in NGC 6752.  Whereas Shetrone was unable to separate the
contribution of $^{25}$Mg from $^{26}$Mg, we will make this important
distinction.
In \S\ref{sec:data} the sample
selection and data acquisition are covered.  In \S\ref{sec:param} the
derivation of the stellar parameters is presented.  The elemental
abundances are determined in \S\ref{sec:abund} and Mg isotopic ratios in
\S\ref{sec:mg}.
The discussion is in \S\ref{sec:discuss} and concluding remarks are
given in \S\ref{sec:conc}.


\section{Data Acquisition}
\label{sec:data}

\subsection{Target selection}

\begin{figure}
\centering
\includegraphics[width=8.5cm]{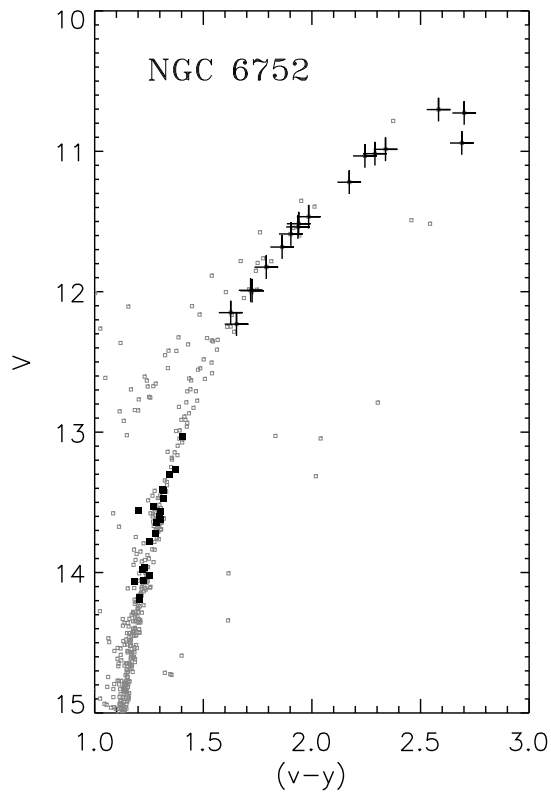}
\caption{The colour--magnitude diagram from NGC 6752.  The
large plus symbols represent the bright red giants reported
here while the filled black squares represent the RGB bump
stars observed under ESO programme 65.L-0165 and reported
by \citet{grundahl02}. \label{fig:cmd}}
\end{figure}

The targets for this study were selected from the $uvby$ photometry 
of NGC 6752 in \citet{grundahl99}.  We selected
the stars such that there would be no detected neighbours within the 
entrance aperture (1.5$\times$2.0 arcsec) of the UVES image slicer in 
order to minimize contamination. 
The cluster colour--magnitude diagram (CMD) shown in Fig.~\ref{fig:cmd} was used 
to select the best first ascent RGB stars.  The sample is 
free from AGB contamination fainter than $V=11.4$.  Brighter than 
this the contamination is very unlikely since the evolutionary 
timescale on the RGB is much slower than for AGB stars.

Three additional stars (B 702, B 708, A88) used in the analysis are not 
shown in Fig.~\ref{fig:cmd}.  All 3 stars were analyzed by \citet{shetrone98}.
Reobservations were made for B 702 and A88 and all three stars were
reanalyzed.  The $V$ magnitudes and $B-V$ colours place these stars 
at the tip of the RGB.

\subsection{Observations and reduction}

The observations for this programme were carried out with the UVES 
instrument on the ESO VLT UT2 telescope using the red arm in the 
standard setting of center 580nm and cross disperser \#3. 
In order to separate the 
Mg isotopic components, the highest possible spectrograph resolution
$R=110,000$ was used for all observations.  This corresponds to a slit
width of 0\farcs3, so we used the image slicer (\#3) to minimize light 
losses due to seeing.  Our programme stars were observed in service
mode during the summer of 2001.  The integration time for each star
was split in three to be able to remove the effects of cosmic ray hits
on the CCD detector and the total exposure times ranged between 3600 and 5400
seconds, depending on the brightness of the target star.  Our spectra
covered the wavelength region between 4735{\AA} and 6830{\AA} with a
small gap between 5800{\AA} and 5825{\AA} due to the space between the
two CCDs in the UVES camera.   The signal-to-noise ranged from 250
to 150 per pixel.
In addition, we also observed the hot star HR6788 in order to use it
for removal of telluric features (using {\tt telluric} in 
IRAF\footnote{IRAF is distributed 
by the National Optical Astronomy Observatories,
which are operated by the Association of Universities for Research
in Astronomy, Inc., under cooperative agreement with the National
Science Foundation.}) near 
the [O{\sc i}] line at 6300{\AA}. 

Standard IRAF procedures were used for bias subtraction, correction for
interorder background, flat fielding and extraction of
spectra.  The total width of the 5 slices in each order was 10 arcsec 
and for each wavelength bin we summed the flux over all 
corresponding pixels.  Wavelength calibration was carried out on a nightly
basis using Th-Ar spectra. 

By using the image slicer in UVES the sky spectrum could not be measured 
simultaneously with the star spectrum.  We note, however, that even for the 
faintest stars in our sample, $V\approx12$, observed in full Moon periods
the sky signal integrated over the 10$\times$0.3 arcsec slit is less than
0.3\% of the star signal in the spectral region from 4800{\AA} to 6800{\AA}.
Hence, the sky contribution can be neglected.

For the three additional stars, A88 and B 708 were observed on the VLT in service mode
during the summer of 2000 with $R=110,000$ and signal-to-noise around 100.  
B 702 was observed with the ESO CAT with $R=80,000$, for details 
about this observation see \citet{shetrone98}.


\section{Stellar parameters}
\label{sec:param}

\subsection{\teff~and log g}

For determining the effective temperature and surface gravity of our target
stars we used the same procedure as described by \citet{grundahl02}, 
which we shall briefly repeat here.  Temperatures were derived from the $uvby$ 
photometry using the $(b-y)$ index and the calibration by \citet{alonso99b}, 
adopting E$(B-V)\,=0.04$ \citep{harris96} for the cluster reddening.  To
estimate the surface gravity we adopted an apparent distance modulus of
$(m-M)_V\,=\,13.30$ for the cluster and a stellar mass of 0.84 $M_\odot$. 
The bolometric corrections were taken from a 14 Gyr isochrone with 
[Fe/H]=$-$1.54 from \citet{vandenberg00}.  The stellar parameters
are presented in Table \ref{tab:abpar}.  We found that changes 
of $\pm0\fm1$ in distance modulus led to changes of
$-$0.04dex in the derived gravity and changes of $\pm0\fm02$ in $(b-y)$ 
(reddening and calibration uncertainty) resulted in errors in the temperature of
approximately 40K (average value for the 17 
stars).  From Table \ref{tab:aberr} we therefore 
see that the uncertainty of distance modulus and cluster photometry have very 
little impact on our derived results.


\subsection{Microturbulence}

For each star, we determined the abundance of key elements beginning with
the iron abundance as the canonical measure of metallicity. A model
atmosphere was taken from the \citet{kurucz93} LTE stellar atmosphere
grid. We interpolated within the grid when necessary to get a model
with the required \teff, log g, and [Fe/H].
The model was used with the LTE stellar line analysis program
{\sc Moog} \citep{moog}. The equivalent width of a line was measured using
IRAF where in general a Gaussian profile was fit to an
observed profile.  To complete the abundance analysis, an estimate of the
microturbulence is needed. This was determined from Fe\,{\sc i} lines.

\begin{figure}
\centering
\includegraphics[width=8.5cm]{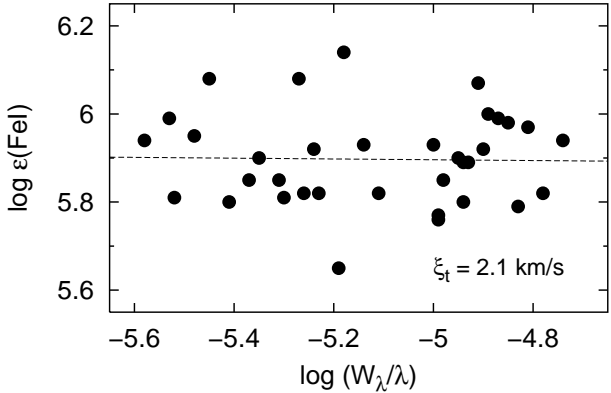}
\caption{The abundances derived from individual Fe\,{\scshape i} lines
plotted against the reduced equivalent width (W$_\lambda/\lambda$) for star NGC6752-mg6.  
The microturbulence, $\xi_t$, is set by the requirement that the abundances (derived 
from individual Fe\,{\scshape i} lines) show no trend against 
(W$_\lambda/\lambda$).  The dotted line represents 
the linear least squares fit to the data. \label{fig:micro}}
\end{figure}

A selection of 31 Fe\,{\sc i} and 7 Fe\,{\sc ii} lines was measured. 
The $gf$-values were taken mostly from \citet{lambert96}
and a small number from a list compiled by R.E. Luck (1993, private
communication).  The microturbulence  ($\xi_t$ in km s$^{-1}$)
was determined in the usual way by insisting that
the Fe abundance from Fe\,{\sc i} lines be independent of their
equivalent width ($W_\lambda$). An example of the abundance versus
$W_\lambda$ plot is shown in Fig.~\ref{fig:micro}. The derived
$\xi_t$ range from about 2.7 km s$^{-1}$ for the stars at the tip of the
giant branch to 1.8 km s$^{-1}$ for our faintest stars which are
1.5 magnitudes below the tip. The measurement uncertainty is typically
about $\pm$ 0.2 km s$^{-1}$.  

\input{3481.t1}


\subsection{The Iron Abundance}

\begin{figure}
\centering
\includegraphics[width=8.5cm]{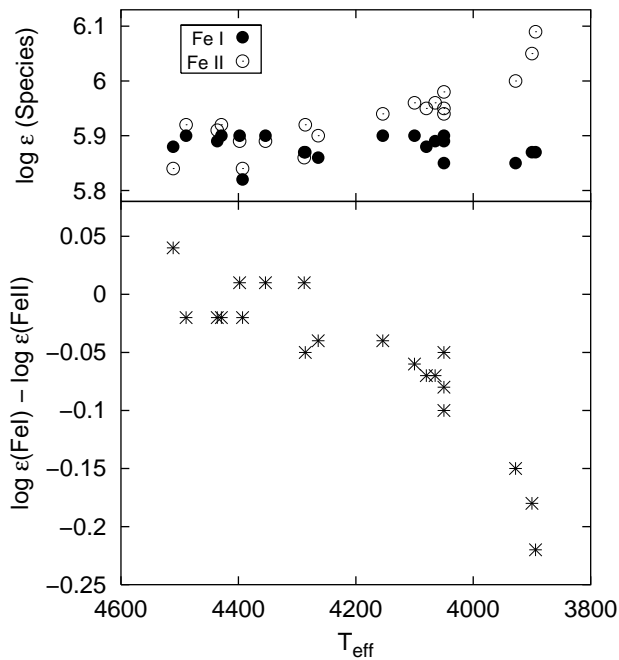}
\caption{In the upper panel we plot log$\epsilon$(Fe{\scshape i}) 
(filled circles) and log$\epsilon$(Fe{\scshape ii}) (open circles) versus \teff.
In the lower panel we plot the difference 
between Fe {\scshape i} and Fe {\scshape ii}
abundance against \teff.  We find that log$\epsilon$(Fe{\scshape i}) remains
essentially constant as \teff~changes while log$\epsilon$Fe({\scshape ii})
increases with decreasing \teff~beginning around \teff=4200K.
\label{fig:fe}}
\end{figure}

The Fe abundances from the Fe\,{\sc i} lines are but weakly dependent on the
derived $\xi_t$ because the set of Fe\,{\sc i} lines contains
an ample number of weak lines.  These abundances show a dispersion attributable
to measurement errors.  The mean abundance is $\log\epsilon$(Fe) =
5.89 $\pm$ 0.01 ($\sigma=0.02$) where the standard deviation is consistent with all
stars having the same abundance.
Determination of the Fe abundance from the smaller set of Fe\,{\sc ii}
lines offers a check on the adopted stellar parameters and assumptions
underlying the model atmospheres and the line analysis. The Fe\,{\sc ii}
and Fe\,{\sc i} lines give the same abundance for the warmer stars
(\teff $\ge 4200$ K) but cooler stars show a discrepancy which
increases with decreasing \teff~(Fig.~\ref{fig:fe}).  The coolest
giants show an Fe abundance about 0.2 dex larger from Fe\,{\sc ii} 
than from  Fe\,{\sc i} lines. 
To place this systematic effect in perspective, we present in Table \ref{tab:aberr}
the customary table showing the effect on the derived Fe abundance from
Fe\,{\sc i} and Fe\,{\sc ii} lines of small changes to the model
atmosphere parameters. By inspection, we see that a combination of
changes to the parameters within their measurement uncertainties would 
allow the neutral and ionized lines to return the same Fe abundance
for the coolest giants. 
For example, a temperature correction running from an increase
of \teff~by 100 K at 3900 K and vanishing at about 4200 K would
remove the discrepancy among the coolest stars and ensure that all
stars gave the same Fe abundance.  This mild revision of the
\teff~scale seems preferable to the supposition that the
discrepancy reflects the appearance of non-LTE (or other) effects in
the atmospheres of the coolest stars. If the Fe abundance is
derived from Fe\,{\sc i} and Fe\,{\sc ii} lines of stars with
\teff$ > 4200$ K, the weighted mean abundance is
$\log\epsilon$(Fe) = 5.88 $\pm$ 0.01 ($\sigma=0.03$).  
Taking the \citet{kurucz84} solar atlas, 
we measured the equivalent widths of Fe\,{\sc i} and Fe\,{\sc ii} lines
(same set of lines as measured in the giants).  The measured equivalent
widths range from 10 to 90 m\AA.  Using a Kurucz model with
\teff=5770K, log g=4.44, $\xi_t$=1.0 km s$^{-1}$, we derive 
log$\epsilon$(Fe\,{\sc i}) = 7.51 and 
log$\epsilon$(Fe\,{\sc ii}) = 7.50 using the Van der Waals line damping
parameter (Uns{\" o}ld approximation multiplied by a factor recommended by the
Blackwell group).  Using the \citet{holweger74}
empirical model atmosphere with $\xi_t$=1.15 km s$^{-1}$, we derive log$\epsilon$(Fe\,{\sc i}) = 7.53 and 
log$\epsilon$(Fe\,{\sc ii}) = 7.47.  
Adopting a solar metallicity
of $\log\epsilon$(Fe) = 7.50, we obtain [Fe/H] = $-$1.62 for
NGC 6752 with no evidence for a star-to-star variation. 

\input{3481.t2}

Our cluster metallicity [Fe/H] = $-$1.62 is similar to previously
published values.  Grundahl et al. (2002, private communication)  
also obtained [Fe/H] = $-$1.62 from high-resolution spectra of
their red giants.  \citet{gratton01} report [Fe/H] = $-$1.42 from
UVES spectra of turn-off stars and subgiants.  \citet{kraft02}
find [Fe/H]$_{\rm I}=-1.50$ and [Fe/H]$_{\rm II}=-1.42$ using
Kurucz models and [Fe/H]$_{\rm I}=-1.51$ and [Fe/H]$_{\rm II}=-1.50$
using MARCS models.  Other analyses of
NGC 6752's giants include [Fe/H] = $-$1.42 \citep{carretta97},
$-$1.58 \citep{minniti93}, and $-$1.54 \citep{zinn84}.  
That we find NGC 6752 to be slightly more metal-poor than
other investigators is not relevant to our main purpose which is
the measurement of the Mg isotopic ratios and the abundances of the
light elements O, Na, Mg, and Al.

For the three additional stars, the VLT reobservations revealed 
no discernable systematic differences in derived abundances 
when compared to the original observations and analysis \citep{shetrone98}.


\section{Oxygen, Sodium, Magnesium and Aluminium Abundances}
\label{sec:abund}
\label{sec:el}

Red giants in NGC 6752 are known to show star-to-star
variations in the C and N abundances. \citet{norris81}
found the variations in CN band strength anticorrelated with
CH band strength.  \citet{ss91}, as noted earlier,
found the star-to-star variations to extend to main sequence
stars. Extension of star-to-star abundance variations to
O, Na, Mg, and Al has been reported for several clusters
(e.g. \citealt{kraft97,sneden97,M5}).
In the case of NGC 6752, this extension has been made 
for a handful of bright giants \citep{norris95,minniti96}, 
red giant branch stars \citep{grundahl02},
and turn-off stars and early subgiants \citep{gratton01}.  Here,
we extend the search for abundance variations to a large sample of stars at
the top of the red giant branch. 

\begin{figure}
\centering
\includegraphics[width=8.5cm]{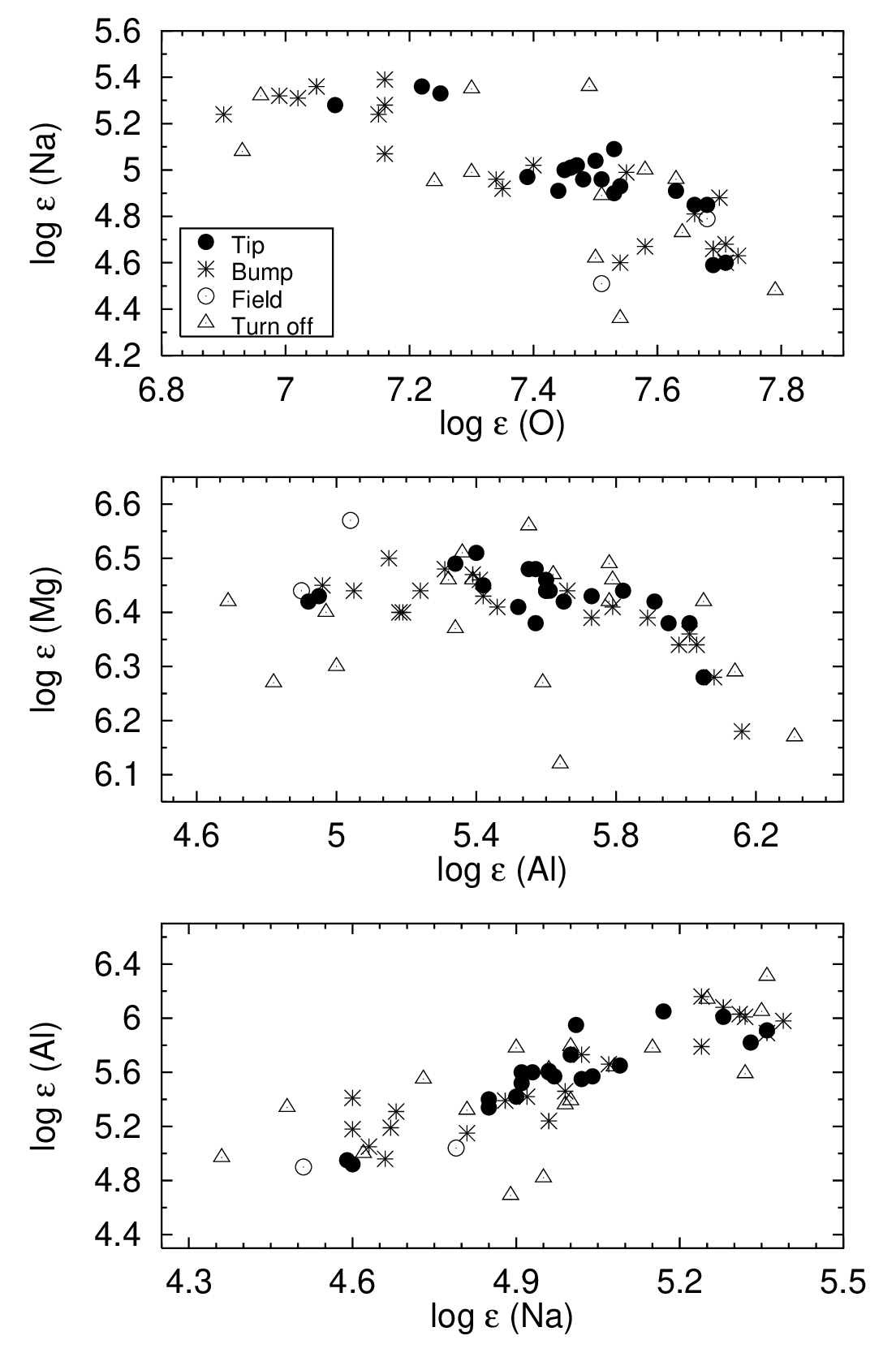}
\caption{ Observed anti-correlations between O and Na (upper), Al and 
Mg (middle), and the correlation between Na and Al (lower).  The 
RGB tip stars from this study (closed circles),
the \citet{grundahl02} RGB bump stars (asterisks), the \citet{gratton01}
turn-off stars and early subgiants (open triangles), and the 2 field
giants (open circles) are plotted.   
Abundances from \citet{gratton01} were shifted onto our scale (see text).
\label{fig:onamgal}}
\end{figure}

For O, Na, Mg, and Al, abundances were derived from the measured
equivalent widths using the $g$f-values from \citet{shetrone96a} 
and from a list compiled by R. E. Luck (1993, private communication).  
The radial velocity of the cluster stars ranged from $-$17 to $-$46 km s$^{-1}$
allowing easy removal of night sky emission without affecting stellar
O absorption lines.  Our line list is
presented in Table \ref{tab:line}.  The elemental abundances are presented
in Table \ref{tab:abpar} and the anticorrelations between O-Na and
Mg-Al can be seen in Fig.~\ref{fig:onamgal}.  As 
seen in previous studies, we find an anticorrelation between
oxygen and sodium where stars with low abundances of O have
correspondingly large abundances of Na.  Within the bright giants,
oxygen varies by $\Delta$log$\epsilon$(O) = 0.6, sodium varies
by $\Delta$log$\epsilon$(Na) = 0.8, aluminium varies 
by $\Delta$log$\epsilon$(Al) = 1.1 while 
magnesium varies by $\Delta$log$\epsilon$(Mg) = 0.2.  There is
a weak anticorrelation between Mg and Al.  

\input{3481.t3}

Taking NGC6752-mg0 and NGC6752-mg24, stars at the
cool and warm end of our sample, we determined the dependence
of the derived abundances upon the model parameters 
(see Table \ref{tab:aberr}).  
Our results show that oxygen decreases by about 0.6 dex as sodium increases
by about 0.8 dex (Fig.~\ref{fig:onamgal}).  These changes in elemental abundance are
far greater than the estimated uncertainties in Table \ref{tab:aberr}. 
The aluminium abundance range is about 1.1 dex, again at least an order of magnitude
larger than the estimated uncertainty.  The small spread of the Mg
abundance across our sample probably exceeds the estimated
uncertainties; note the sensitivity of the Mg abundance to the
microturbulence (Table \ref{tab:aberr}). The Na and Al abundances are well
correlated.

In order to compare our light element abundances with those
for less luminous red giants, we have
reanalyzed Grundahl et al.'s $W_\lambda$s for 
lines in our line list using Kurucz models
for the atmospheric parameters found by them.  Results plotted
in Fig.~\ref{fig:onamgal} show excellent agreement for O, Na, Mg, and Al 
abundances with respect to absolute abundance and total spread.  
Grundahl et al.'s sample appears to extend the Na-O anticorrelation at the
high-Na low-O end. 
If we use the MARCS model atmospheres used by
Grundahl et al., their abundances are typically about 0.15 dex lower.
Regarding the Mg-Al anticorrelation, we note that 
for both our sample and the Grundahl et al.\ sample, the total
$\log\epsilon$(Mg) + $\log\epsilon$(Al) is constant.  This is 
consistent with previous work by \citet{shetrone96a,shetrone98} for M 13
and NGC 6752.

A comparison is possible with results for turn-off
stars and subgiants using [O/Fe], [Na/Fe], [Mg/Fe], and
[Al/Fe] published by \citet{gratton01}.  Their O 
abundances are derived from the 7771-7774 \AA\ triplet;
non-LTE effects were considered but are 0.1 dex or smaller.
The Na abundance is based on the `quite strong' 8183\AA\ and
8194 \AA\ doublet and includes a small non-LTE correction.
Converting published results to abundances $\log\epsilon$(Species),
we find that their Na-O anticorrelation parallels 
ours.\footnote{In this conversion, we assume 
solar abundances of 6.33 for Na and 8.92 for O, and
[Fe/H] = $-$1.42, Gratton et al.'s value.}  Offsets to
Gratton et al.'s abundances of $-$0.2 dex for Na and $-$0.2 dex
for O superimpose their results on ours (see Fig.~\ref{fig:onamgal}).

Gratton et al.'s Al abundances for subgiants are derived from
8774 \AA\ doublet and include a small non-LTE correction. For the
turn-off stars, the Al abundance depends on the 3944 \AA\ and
3961 \AA\ resonance lines and a large non-LTE correction. The
Mg abundance was found from a set of lines in the
blue and yellow regions of the spectrum that likely includes lines
from our line list.  
The range in Al abundances\footnote{Conversion of [Al/Fe] to Al abundance was
made assuming a solar Al abundance of 6.47 and [Fe/H] = $-$1.42.} 
from Gratton et al.'s sample
runs from about 4.6 to 5.8, which is displaced to lower
abundances by about 0.4 dex relative to our range.
Gratton et al.'s Mg abundances are systematically lower than
ours. Their result [Mg/Fe] $\simeq 0.0$ contrasts with
[Mg/Fe] $\simeq$ 0.6 for our and Grundahl et al.'s
samples of giants. 
Offsets to Gratton et al.'s abundances of $+$0.2 dex for 
Mg\footnote{Conversion of [Mg/Fe] to Mg abundance was
made assuming a solar Mg abundance of 7.58 and [Fe/H] = $-$1.42.} and $+$0.4 dex for Al 
superimpose their results on ours (see Fig.~\ref{fig:onamgal}).  Clearly, 
it would be helpful to understand the origins of the
differences in the Mg and Al abundances.  In particular, interpretations
of star-to-star abundance variations are severely constrained if the differences
are real and not a reflection of systematic errors in one or both of the
abundance analyses.

Our stellar sample includes two field stars analyzed by \citet{shetrone96b}.  The
O, Na, Mg, and Al abundances obtained from our $W_\lambda$s and a Kurucz model for
Shetrone's chosen \teff~and log g are given in Table \ref{tab:abpar}.  We add 
these stars to Fig.~\ref{fig:onamgal} by adjusting the abundances for the small
metallicity differences between the stars and NGC 6752 by supposing
that $\Delta\log\epsilon$(Species) = $\Delta\log\epsilon$(Fe).  It is to be noted
that the comparison stars fall at the high O, low Na, high Mg, and low Al end of the
distribution defined by the red giants.  Our Fe, O, Na, Mg, and Al abundances
agree with Shetrone's values to better than 0.1 dex with just two minor
exceptions: our O abundance for HD 141531 is 0.14 dex larger and the Al
abundance for HD 103036 is 0.2 dex larger than Shetrone's values.


\section{Magnesium Isotopic Abundances}
\label{sec:mg}

\begin{figure}
\centering
\includegraphics[width=8.5cm]{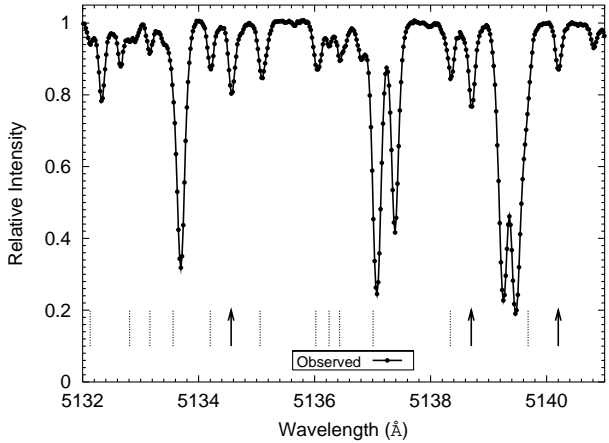}
\caption{Spectra of NGC6752-mg6 from 5132 to 5141 \AA.  The positions of
the MgH A-X 0-0 and MgH A-X 1-1 lines are marked below the spectrum.  The 
majority of MgH lines are unsuitable
for isotopic analysis due to blends.  The positions 
of the 3 features that we used to derive
the isotopic ratios are highlighted with arrows. \label{fig:mghregion}}
\end{figure}

\begin{figure}
\centering
\includegraphics[width=8.5cm]{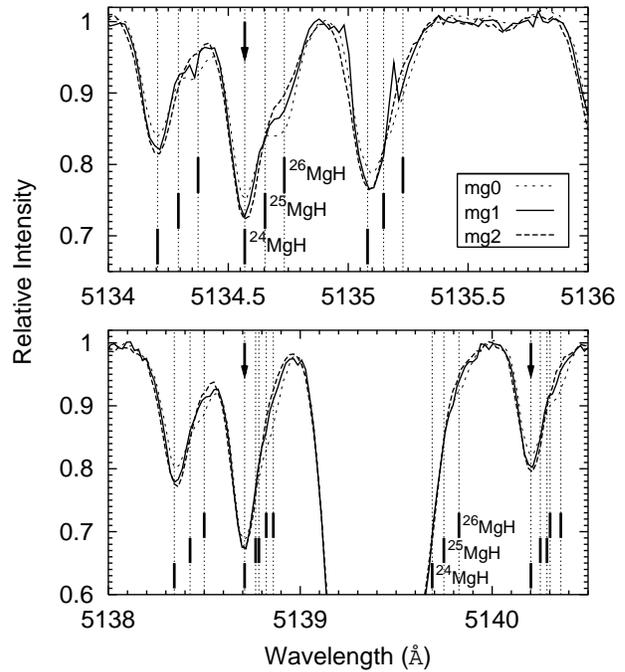}
\caption{Spectra of NGC6752-mg0, mg1, and mg2 from 5134 to 5136 \AA~(upper)
and from 5138 to 5140.5 \AA~(lower).  The positions of
the $^{24}$MgH, $^{25}$MgH, and $^{26}$MgH lines are shown.  The
lines used to derive the Mg isotopic ratios are identified by
arrows.  Striking differences in the line profiles are 
immediately obvious where $^{26}$MgH appears to be varying
between the 3 stars. \label{fig:mg012}}
\end{figure}

The A-X band of MgH is present in the spectra
of NGC 6752's red giants.  Shetrone's pioneering study of Mg isotopic
abundances in red giants from M 13, a cluster of similar
metallicity with similarly pronounced O, Na, Mg, and Al abundance
variations, demonstrated the feasibility of the program to determine the
isotopic abundances.  \citet{shetrone98} also determined
Mg isotopic ratios for 4 bright giants in NGC 6752.
Examination of the spectra confirms that MgH lines are
present - see Fig.~\ref{fig:mghregion}.  More importantly, a mere glance at
the spectra shows that there is a star-to-star variation in the
isotopic abundance ratios.  This is well shown by Fig.~\ref{fig:mg012}
where we overplot the spectra of the three coolest (brightest)
giants in the regions providing three of the key MgH lines selected
by \citet{ml88} as indicators of the isotopic
abundance ratios.  Note the asymmetric profiles of the MgH lines
with the $^{25}$MgH and $^{26}$MgH lines providing the
trailing red wing.  For each of the three lines (and others), the
relative strengths of the asymmetry are the same with the strongest 
asymmetry occurring for mg0 and the weakest for mg2. Inspection of
Fig.~\ref{fig:mg012} also shows that the differences in the profile occur at the
wavelength of the $^{26}$MgH components but appear to vanish at the
wavelength of the $^{25}$MgH components; the abundance differences
are largely in the $^{26}$Mg/$^{24}$Mg ratio not in 
the $^{25}$Mg/$^{24}$Mg
ratio.  To convert these impressions to quantitative estimates,
spectrum syntheses were generated and fitted to the observed
spectra. 

\begin{figure}
\centering
\includegraphics[width=8.5cm]{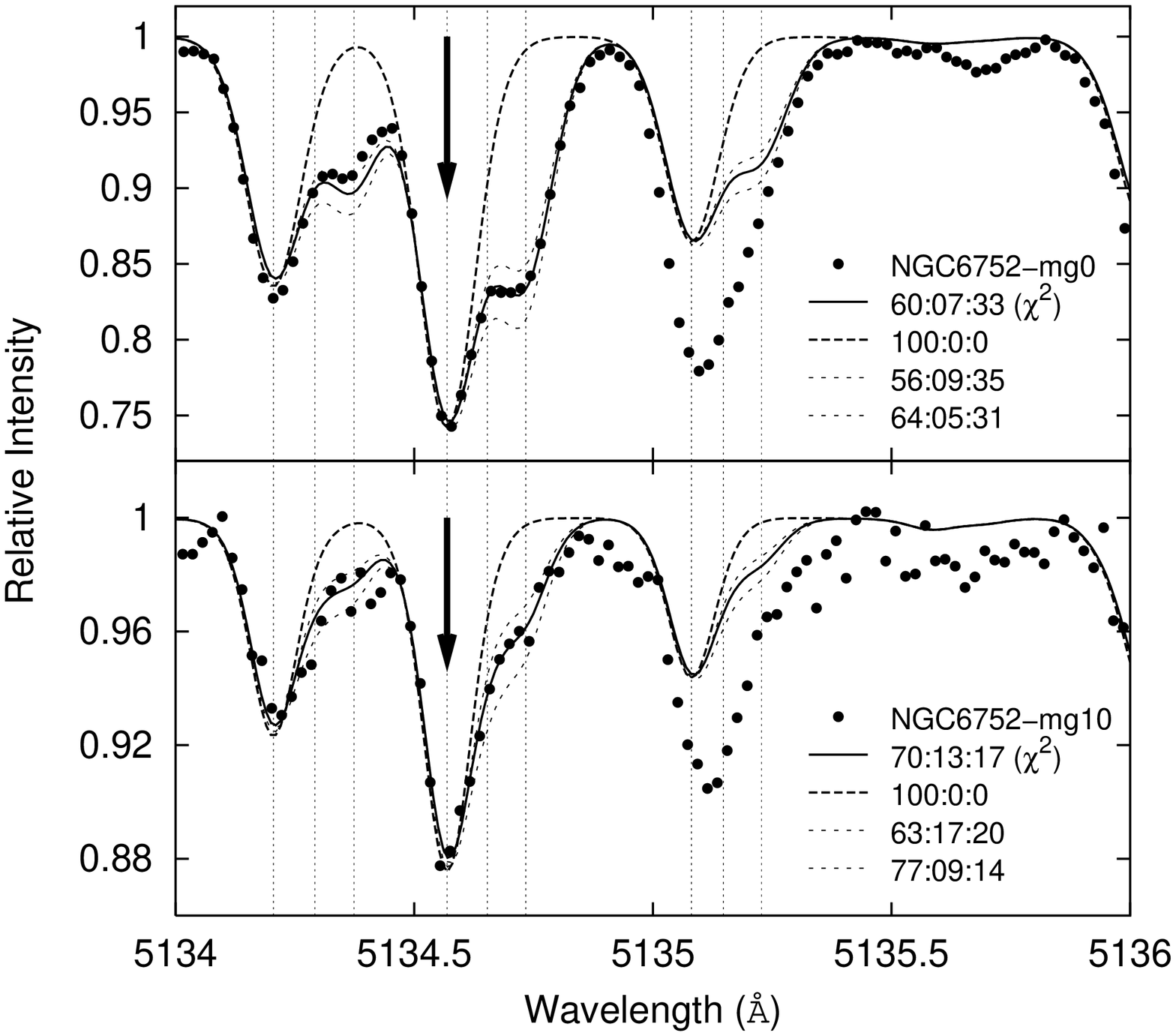}
\caption{Spectra of NGC6752-mg0 (upper) and NGC6752-mg10 (lower) from 
5134.0 to 5136.0 \AA.  The feature we are interested in fitting
is highlighted by the arrow.  The positions of the $^{24}$MgH, $^{25}$MgH, 
and $^{26}$MgH lines are indicated by dashed lines.  The 
closed circles represent the observed spectra.  
The synthetic spectrum generated using the isotopic ratios
determined by $\chi^2$ analysis is given by the solid line: the
$^{24}$Mg:$^{25}$Mg:$^{26}$Mg ratios are given on the figure.  
Unsatisfactory ratios are plotted as dotted lines.
\label{fig:region1}}
\end{figure}

In the quantitative analysis, three MgH features
recommended by \citet{ml88} and used by
\citet{gl2000} were analyzed.
The first line at 5134.6 \AA~is a blend of the $Q_1(23)$ and $R_2(11)$
lines from the 0-0 band.  There are 2 similar features on either
side of this line where contributions from other lines render them unsuitable for
the extraction of reliable isotopic ratios \citep{tl80}.  The second line at
5138.7 \AA~is a blend of the 0-0 $Q_1(22)$ and 1-1 $Q_2(14)$ lines
in the wing of a strong atomic line.  The last line employed in the
determination of the isotopic ratios is at 5140.2 \AA, a blend
of the 0-0 $R_1(10)$ and 1-1 $R_2(4)$ lines.
The macroturbulence was determined with a Gaussian representing the combined
effect of the atmospheric turbulence, stellar rotation, and the
instrumental profile.  This Gaussian was fixed by fitting the profiles of
unblended lines namely the Ni {\scshape i} 5115.4 \AA~and the
Ti {\scshape i} 5145.5 \AA.  In all stars, these lines were slightly stronger than
the recommended MgH lines.  The line list was identical to that
used in \citet{gl2000} and included the following elements
C, Mg, Sc, Ti, Cr, Fe, Co, Ni, and Y.  The $^{25}$Mg and $^{26}$Mg
abundances were adjusted by trial and error until the profiles
of the 3 recommended features were best fitted where we treated each
of the 3 regions independently.  The best
fit was determined by eye (see Table \ref{tab:iso} for Mg isotopic ratios).  
Examples of spectrum syntheses are shown in Fig.~\ref{fig:region1}.  A point
deserving of emphasis is that the red asymmetry of MgH lines demands major
contributions from the $^{25}$Mg and especially the $^{26}$Mg isotopes.  A
spectrum computed for pure $^{24}$Mg clearly does not fit the observed spectrum.
We refrain
from using the MgH features to determine the Mg abundance as slight
variations in the adopted temperature result in large changes to the
Mg abundance required to fit the molecular lines.

\input{3481.t4}

\begin{figure}
\centering
\includegraphics[width=8.5cm]{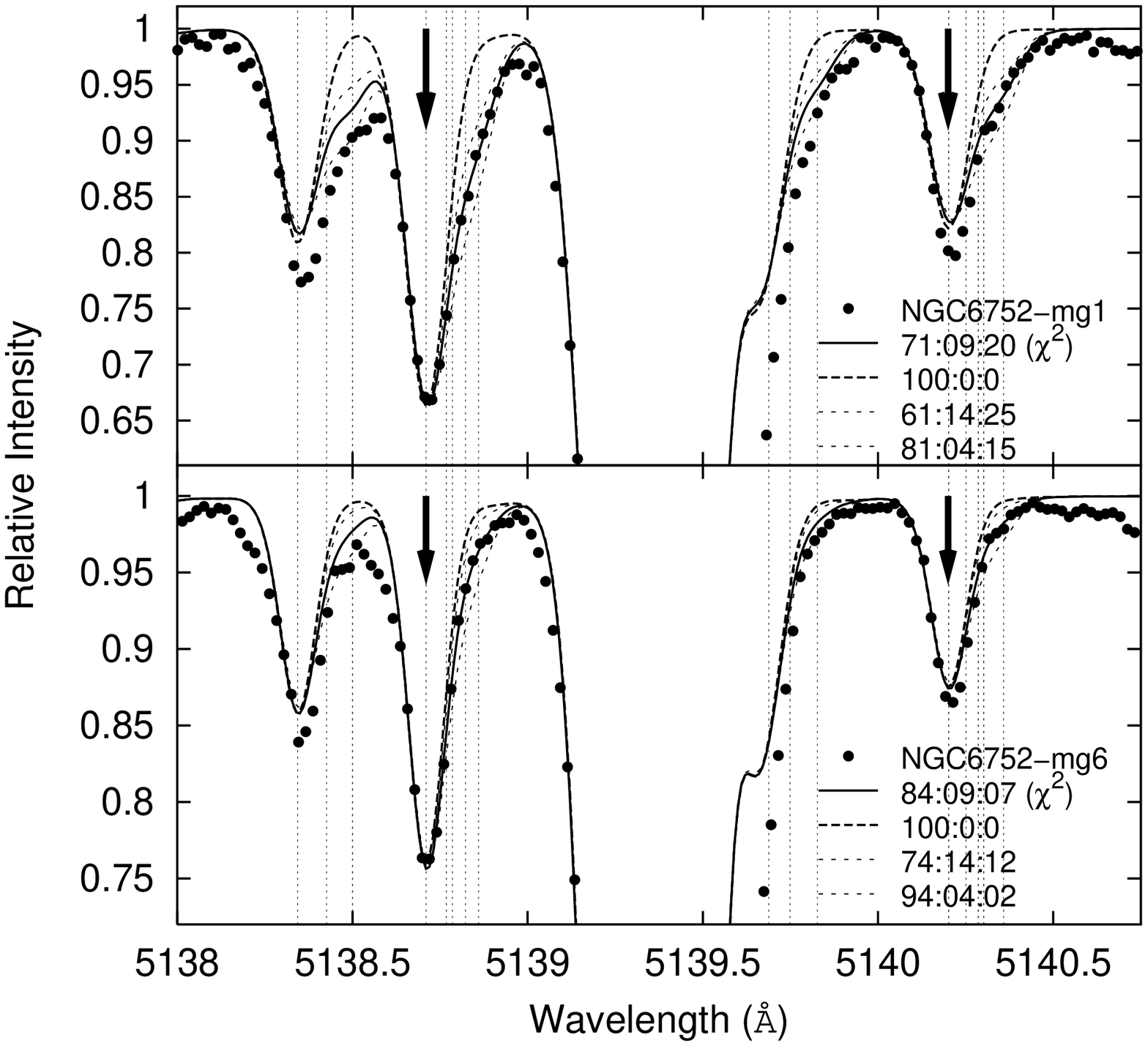}
\caption{Spectra of NGC6752-mg1 (upper) and NGC6752-mg6 (lower) for 
5138.0 to 5140.8 \AA.  The features we are interested in fitting
are highlighted by the arrows.  The positions of the $^{24}$MgH, $^{25}$MgH, 
and $^{26}$MgH lines are indicated by dashed lines.  The 
closed circles represent the observed spectra.  
The synthetic spectrum generated using the isotopic ratios 
determined by $\chi^2$ analysis is given by the solid line: the 
$^{24}$Mg:$^{25}$Mg:$^{26}$Mg ratios are given on the figure.  
Unsatisfactory ratios are plotted as dotted lines.  
\label{fig:region23}}
\end{figure}

\begin{figure}
\centering
\includegraphics[width=8.5cm]{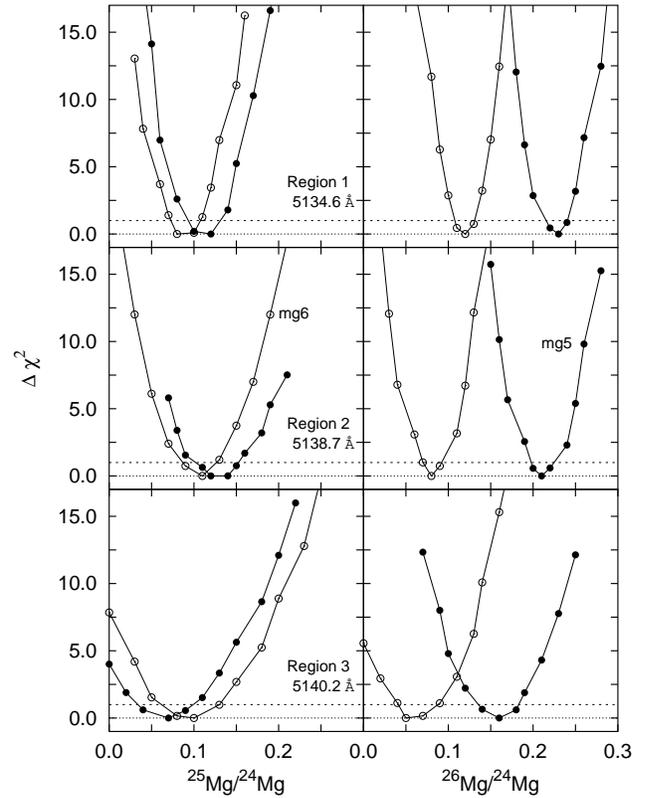}
\caption{Variation of the $\chi^2$ fit for NGC6752-mg5 (closed circles) 
and NGC6752-mg6 (open circles).  The left panels
show the variation of $\chi^2$ for various values of
$^{25}$Mg/$^{24}$Mg while the right panels show the
variation of $\chi^2$ for various values of
$^{26}$Mg/$^{24}$Mg.  The upper panels show the
variation of $\chi^2$ for Region 1 (5134.6 \AA),
the middle panels represents Region 2 (5138.7 \AA), and
the lower panels represents Region 3 (5140.2 \AA). 
The line indicating 1$\sigma$ ($\Delta \chi^2 = 1$) errors
is shown.  
\label{fig:chi.mg26}}
\end{figure}

Next, we sought an unbiased method for determining the best fit to the 
data.  Following the successful work by
\citet{nissen99,nissen00} on the Li isotopic ratios, we
chose to use a $\chi^2$ analysis.  Such an analysis has the benefits
of being unbiased as well as allowing us to quantify the
errors in the fits.  Our free parameters to 
be varied were (1) $^{25}$Mg/$^{24}$Mg,
(2) $^{26}$Mg/$^{24}$Mg, and (3) log$\epsilon$(Mg) where we treated
each of the three recommended features independently.
This choice of free parameters ensured that equal step sizes 
produced comparable changes in the synthetic spectra.  
Our initial guesses for the isotopic ratios were the
optimum values found using the traditional method.  We explored a 
large parameter space around our initial guesses.  The $\chi^2$ was 
calculated via $\chi^2 = \Sigma(O_i-S_i)^2/\sigma^2$ where
$O_i$ is the observed spectrum point, $S_i$ is the synthesis,
and $\sigma = (S/N)^{-1}$.  We determined the optimum
values for $^{25}$Mg/$^{24}$Mg, $^{26}$Mg/$^{24}$Mg,
and log$\epsilon$(Mg) by finding the minima in $\chi^2$.
We then searched a smaller range in parameter space with finer grid spacing
centered upon the optimum values.  We again measured $\chi^2$
locating the minima for $^{25}$Mg/$^{24}$Mg, $^{26}$Mg/$^{24}$Mg,
and log$\epsilon$(Mg) in each region, in each star.  Therefore, for
a given star, we obtained 3 optimal values for 
$^{25}$Mg/$^{24}$Mg and $^{26}$Mg/$^{24}$Mg from which we readily
recovered the ratio $^{24}$Mg:$^{25}$Mg:$^{26}$Mg (see Table \ref{tab:iso} 
and Figs.~\ref{fig:region1} and \ref{fig:region23}).
In finding the optimal values we generated and tested over 1,500 synthetic spectra
per region.  The minimum $\chi^2_{red}=\chi^2/\nu$, where $\nu$ is the number
of degrees of freedom in the fit, was sufficiently close to 1.  
However, since we were interested in a single isotopic ratio 
for a given star, first we needed to quantify the
errors in the ratios $^{25}$Mg/$^{24}$Mg and $^{26}$Mg/$^{24}$Mg for
each region.  Following \citet{bevington92} and \citet{nissen99,nissen00}, 
we plotted $\Delta\chi^2 = \chi^2 - \chi^2_{min}$ against the 
ratios $^{25}$Mg/$^{24}$Mg and $^{26}$Mg/$^{24}$Mg 
(see Fig.\ref{fig:chi.mg26}).  We took
$\Delta\chi^2 = 1$ to be the 1$\sigma$ confidence limit 
for determining $^{25}$Mg/$^{24}$Mg or $^{26}$Mg/$^{24}$Mg.
Thus for each region of each star, we paired an uncertainty
to the optimized value for $^{25}$Mg/$^{24}$Mg or $^{26}$Mg/$^{24}$Mg.
A weighted mean was calculated giving a single
value of $^{24}$Mg:$^{25}$Mg:$^{26}$Mg for each star (see Table \ref{tab:iso}).

\begin{figure}
\centering
\includegraphics[width=8.5cm]{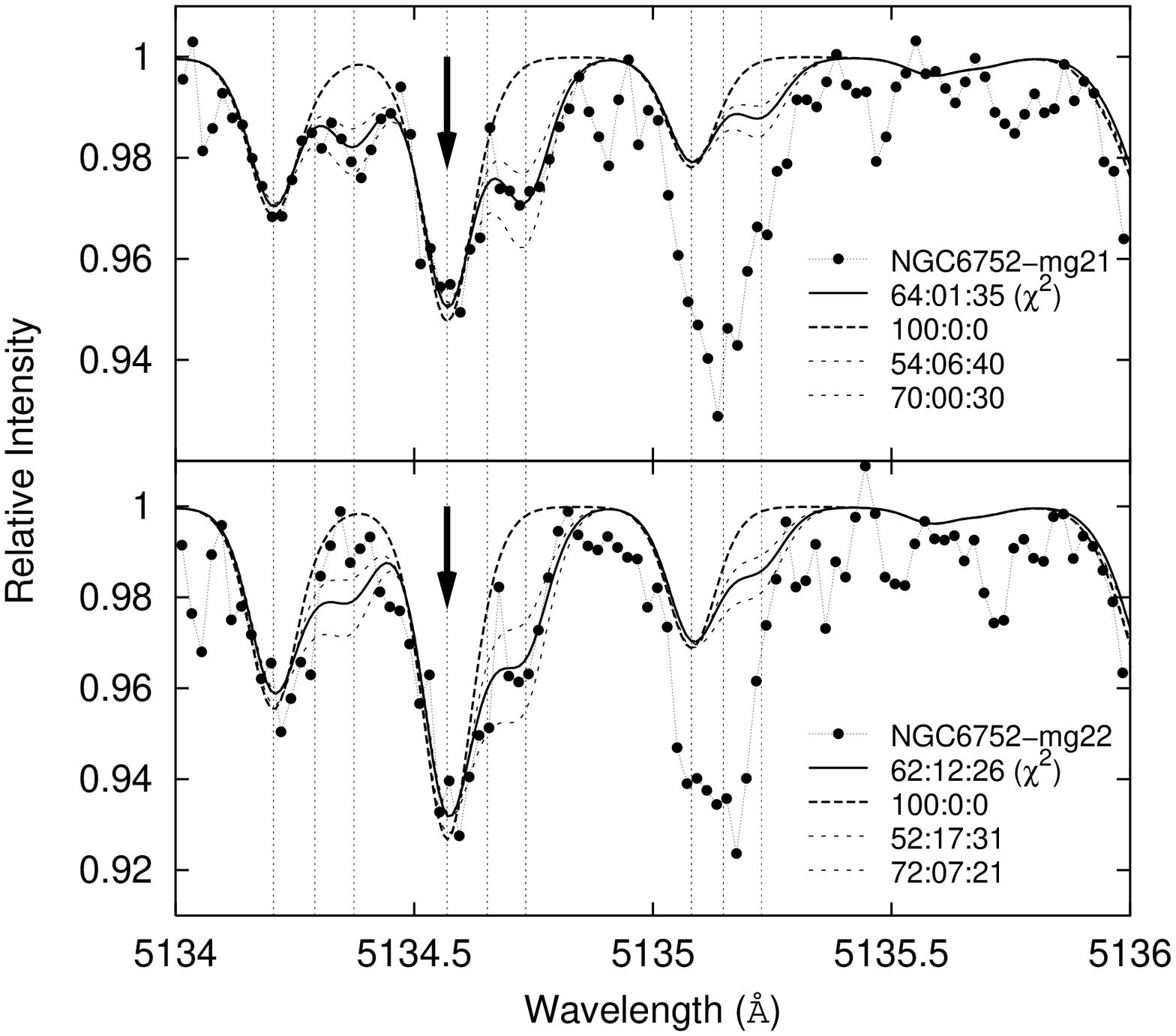}
\caption{Spectra of NGC6752-mg21 (upper) and NGC6752-mg22 (lower) from
5134 to 5136 \AA~showing the MgH lines of interest.  The closed
circles represent the observed spectra.
Although rather weak, the MgH lines show remarkable
asymmetries.  In both panels we plot the best fit (solid line) along
with unsatisfactory ratios (dotted lines).
The asymmetries seen in the MgH lines are not present in atomic lines.
\label{fig:mg2122.1}}
\end{figure}

\begin{figure}
\centering
\includegraphics[width=8.5cm]{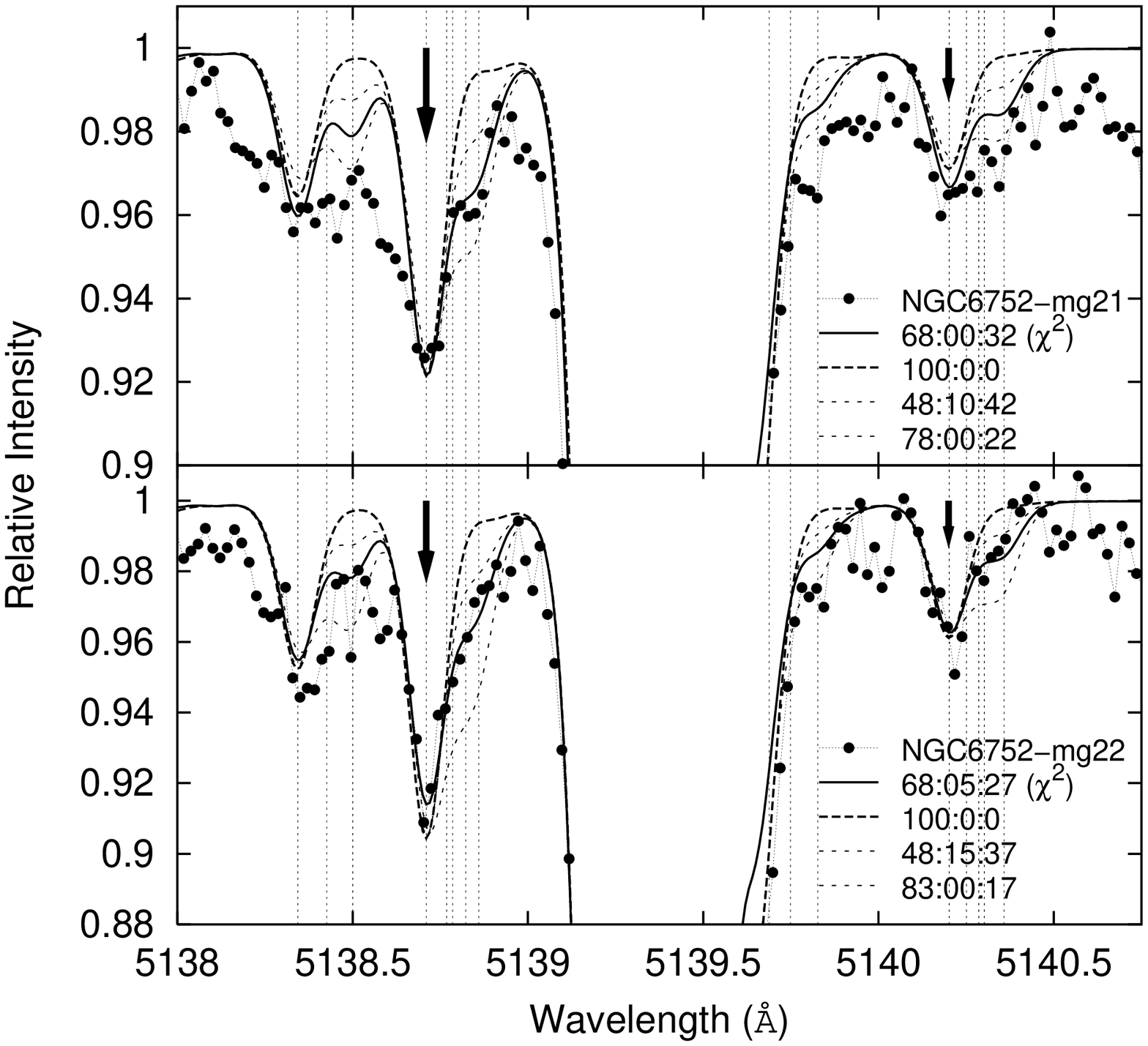}
\caption{Spectra of NGC6752-mg21 (upper) and NGC6752-mg22 (lower) from
5138 to 5141 \AA~showing the MgH lines of interest.  The closed
circles represent the observed spectra.
In both panels, we plot the best fit (solid line) along 
with unsatisfactory ratios (dotted lines).
\label{fig:mg2122.2}}
\end{figure}

We comment on the accuracy of 
the $^{25}$Mg/$^{24}$Mg and $^{26}$Mg/$^{24}$Mg
ratios.  As stated in \citet{gl2000}, the ratio $^{26}$Mg/$^{24}$Mg is
more accurately determined than $^{25}$Mg/$^{24}$Mg due to the larger
isotopic shift.  That is, $^{26}$MgH is less blended with the strong
$^{24}$MgH line than $^{25}$MgH.  From Fig.\ref{fig:chi.mg26}, indeed 
we quantitatively verify that $^{26}$Mg/$^{24}$Mg is
more accurately determined than $^{25}$Mg/$^{24}$Mg.  
Also, we see that in general the ratios
derived using Region 1 are more accurate than those derived using
Regions 2 or 3.  Region 1 has a 
tendency to give higher values of $^{26}$Mg/$^{24}$Mg than 
those obtained from regions 2 or 3.  The dwarfs analyzed
by \citet{gl2000} did not show this trend.  It is likely 
that these highly evolved giants are
affected by unidentified lines which do not affect the dwarfs to the 
same degree.  Table \ref{tab:iso} suggests that the ratios derived
using the traditional approach, where the eye determines the best fit,
compare very favourably with ratios derived using the unbiased $\chi^2$
test.  We now comment upon 2 stars which show unusual MgH lines,
NGC6752-mg21 and NGC6752-mg22.  Figures~\ref{fig:mg2122.1} and 
\ref{fig:mg2122.2} highlight the remarkable asymmetries in the
MgH lines.  The atomic lines in these 2 stars are symmetric.   Though the 
MgH lines are weak, our analyses suggest that both stars have 
$^{26}$Mg/Mg $\ge$ 0.25, a value comparable to 
NGC6752-mg0 and NGC6752-702.

We derived a mean isotopic ratio for each star based on the isotopic
ratios of the 3 regions and their associated errors.  The statistical 
uncertainties in the final ratios were also calculated, though we now comment
on their relevance.  Our formal statistical errors on the mean of the Mg isotopic
ratios are small and fail to take into account systematic errors from various sources.  
Sources of uncertainties in the Mg isotopic ratios include continuum fitting, 
microturbulence, macroturbulence, identified and unidentified blends.  
Errors in the model parameters would equally affect the $^{24}$MgH, $^{25}$MgH, and 
$^{26}$MgH lines and so the isotopic ratio is quite insensitive to the 
selected model parameters.  Inspection of Figs.~\ref{fig:region1}, \ref{fig:region23}, 
\ref{fig:mg2122.1}, and \ref{fig:mg2122.2} show that 
even with exquisite signal-to-noise, it can be difficult to discern by eye
differences in the syntheses below the level b $\pm$ 5 or c $\pm$ 5 when
expressing the ratio as $^{24}$Mg:$^{25}$Mg:$^{26}$Mg=(100$-$b$-$c):b:c.

The three stars observed separately from the rest of the sample (702, 708, A88)
had lower signal-to-noise.  The effect of this lower signal-to-noise can be
estimated by considering NGC6752-mg3.  This star was observed with the main sample
and also with the three stars.  Analysis of the spectrum with inferior signal-to-noise 
gave an isotopic ratio 79:6:15 which compares
favourably with the analysis of the high quality spectrum, 76:9:15.

We focus now upon the two comparison field stars analyzed \citet{shetrone96b},
HD 103036 and HD 141531.  
Shetrone's Mg isotopic ratios for HD 103036 and HD 141531
are 94:03:03 and 90:05:05 respectively while we derive 94:00:06 and
91:02:06 respectively.  Shetrone's data does not allow a distinction
between $^{25}$Mg and $^{26}$Mg and is based upon different MgH lines.
Our data is of higher quality (resolving power and signal-to-noise), we 
separate the contributions of $^{25}$Mg from 
$^{26}$Mg, and rely upon the 3 recommended features used in the
\citet{gl2000} analysis. 


\section{Magnesium Isotopes and Light Element Abundances}
\label{sec:discuss}

\subsection{Observed Correlations and Implications}
\label{sec:obs}

\begin{figure}
\centering
\includegraphics[width=8.5cm]{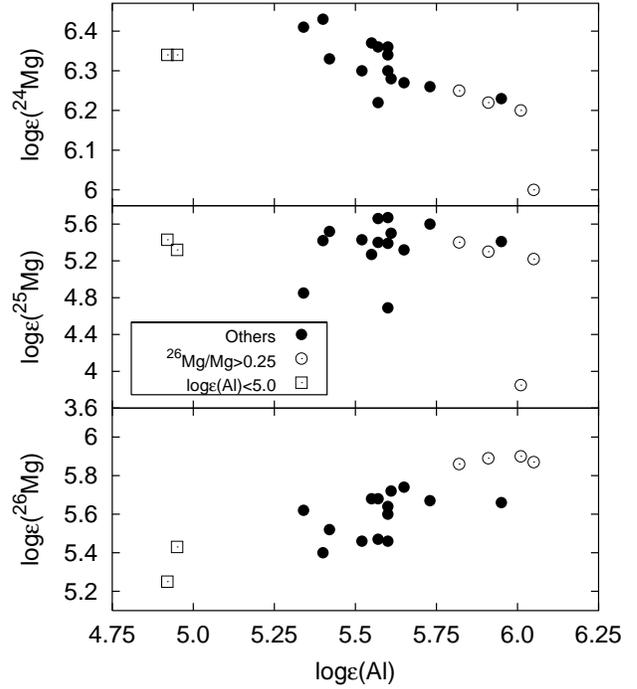}
\caption{The abundances of $^{24}$Mg, $^{25}$Mg, and $^{26}$Mg versus
Al.  The open circles represent the 4 stars with $^{26}$Mg/Mg $>$ 0.25
(NGC6752-mg0,21,22,702), the open boxes represent stars with $\log\epsilon$(Al)$<$5.0,
and the closed circles are the remaining stars.  
There is an anticorrelation of $^{24}$Mg with Al and a correlation of 
$^{26}$Mg with Al.  \label{fig:isovsal}}
\end{figure}

Correlations between O, Na, Mg and Al were
presented in Fig.~\ref{fig:onamgal}.  Here we extend
the search to correlations between the Mg isotopic
abundances and the light element abundances. 
Figure~\ref{fig:isovsal} shows the Mg isotopic abundances and their relation to the
Al abundance.  The isotopic abundances are obtained by combining
the Mg elemental abundance from the Mg\,{\sc i} lines with the
isotopic ratios from the MgH lines.  Isotopic wavelength shifts for
the Mg\,{\sc i} and Al\,{\sc i} lines are so small that these 
provide the respective elemental abundances.  We note in anticipation
of discussion of the nucleosynthesis proposals that this means
that the Al abundance may be the sum of the stable isotope $^{27}$Al and
the radioactive isotope $^{26}$Al with a half-life of 7 $\times$
10$^5$ years.

It is clear that the $^{26}$Mg abundance is well correlated with the
Al abundance (Fig.~\ref{fig:isovsal} - bottom panel).  The total spread in the
$^{26}$Mg abundance is about a factor of 4, a range much greater than
the measurement uncertainties.  In contrast, the $^{25}$Mg abundance
appears to be constant over the 1.1 dex range in the Al abundance.
The $^{24}$Mg abundance declines slightly with increasing Al
abundance.  Our results are consistent with those reported by \citet{shetrone96b} for M13
and by \citet{shetrone98} for NGC 6752.  Shetrone found large amounts of
$^{25}$Mg + $^{26}$Mg relative to $^{24}$Mg in stars with the highest Al and lowest
Mg abundances.  The abundance of $^{24}$Mg decreased as Al increased while
$^{25}$Mg + $^{26}$Mg was not correlated with Al.  Unfortunately, Shetrone's spectra
were not of sufficient quality to allow for a distinction between $^{25}$Mg and $^{26}$Mg.

\begin{figure}
\centering
\includegraphics[width=8.5cm]{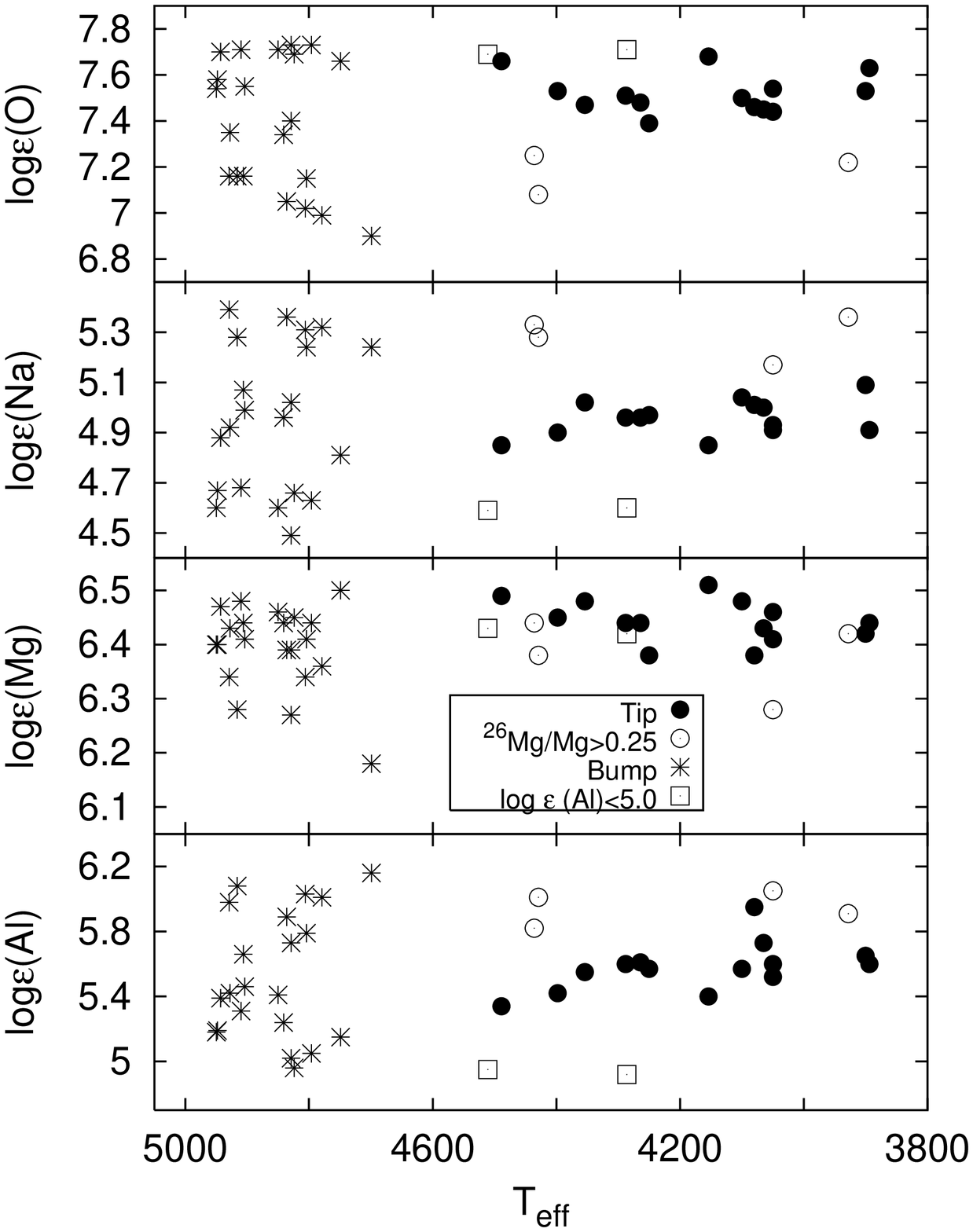}
\caption{The elemental abundances of O, Na, Mg, and Al versus
\teff.  The closed circles represent RGB tip stars from this study, 
the asterisks represent the
\citet{grundahl02} RGB bump stars, the open squares represent the
2 stars with $\log\epsilon$(Al)$<$5.0, and the open circles 
represent the four RGB tip stars with $^{26}$Mg/Mg $>$ 0.25
(NGC6752-mg0,21,22,702).  
Here we take \teff~as a measure of evolutionary status.
None of the elemental abundances exhibit a significant 
correlation with evolutionary status.
\label{fig:elvsteff}}
\end{figure}

\begin{figure}
\centering
\includegraphics[width=8.5cm]{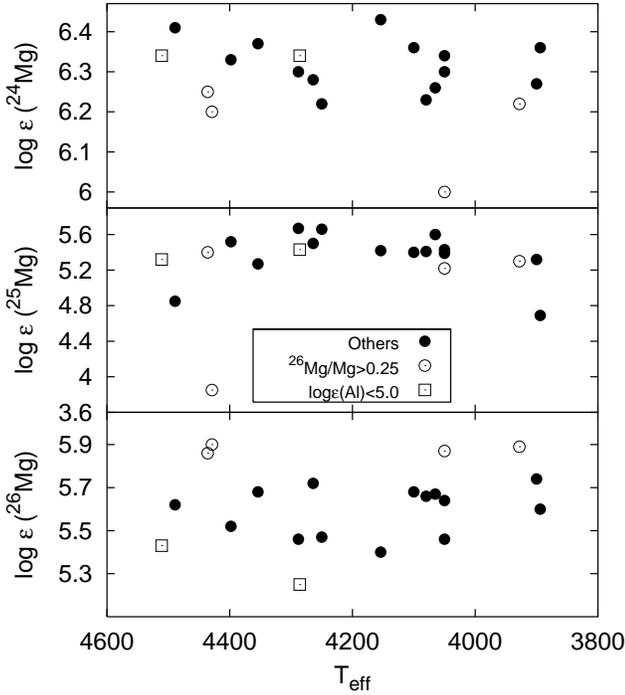}
\caption{The abundances of $^{24}$Mg, $^{25}$Mg, and $^{26}$Mg versus
\teff.  As in figure \ref{fig:elvsteff}, we take \teff~as a measure of 
evolutionary status, the open squares represent the
2 stars with $\log\epsilon$(Al)$<$5.0, and the open circles represent the four stars with 
$^{26}$Mg/Mg $>$ 0.25 (NGC6752-mg0,mg21,mg22).
There are no correlations of the Mg isotopic abundances with evolutionary status.
\label{fig:isovsteff}}
\end{figure}

In the evolutionary versus primordial debate on the origins of the star-to-star
abundance variations, the presence or absence of a correlation of an abundance
with a red giant's luminosity may be a powerful debating point.  Fig.~\ref{fig:elvsteff}
plots the O, Na, Mg, and Al abundances versus \teff, a surrogate for luminosity
for the red giants.  The relation between the Mg isotopic abundances and \teff~is
shown in Fig.~\ref{fig:isovsteff}.  In each panel, our red giants are represented
by one of the three different symbols: the open circles identify the four stars
with the highest $^{26}$Mg abundance, the open squares the two stars with the lowest
Al abundance, the filled circles all other stars.  The open circles and open
squares represent stars at opposite ends of the abundance variation.
We also show on 
Fig.~\ref{fig:elvsteff} results for the \citet{grundahl02} sample of
lower luminosity red giants.  As noted earlier, the ranges of the elemental
abundances agree well with those for our more luminous giants.
There is no evidence in Figs.~\ref{fig:elvsteff} and \ref{fig:isovsteff}
for a correlation between an abundance and luminosity (i.e., \teff).  In particular,
the four $^{26}$Mg richest giants occur at the extremities of the 
\teff-range, and the least luminous stars of our sample present the greatest
spread in abundances.  Grundahl et al.'s collection of stars may be subdivided
at the bump on the red giant branch.  Stars less luminous than the bump show
lithium in their atmospheres but stars more luminous than the bump do not;
shallow convective mixing dilutes lithium by at least a factor of 10 from a maximum
pre-bump abundance of $\log\epsilon$(Li) $\sim 1.0$.  (Note: we use the term
`shallow convective mixing' to refer to mixing responsible for changing C, N, and
Li abundances and `deep mixing' to refer to the mixing responsible for changing the
O, Na, Mg, and Al abundances.)  The abundance
variations (O, Na, Mg, and Al) are the same in giants 
more or less luminous than the bump.  This result
shows that the abundance variations do not reside in a thin layer of accreted material
but are present throughout all of the material swallowed by the convective envelope.

Figs.~\ref{fig:elvsteff} and \ref{fig:isovsteff} do not support a view
that the spread in O-Na-Mg-Al and Mg isotopic abundances arise primarily from internal
processing and deep mixing with these giants.  In contrast to the luminosity
dependent C abundance \citep{ss91}, there is no systematic dependence in
these Figures for any abundance on luminosity (i.e.\ \teff).  Anticipating
the remarks on the dredge-up of H-burning products, the decline of the
C abundance with increasing luminosity should, if deep mixing is
effective, be accompanied by a decline of the O and Mg abundances and an 
increase of the Na and Al abundances.  This is not seen in Fig.~\ref{fig:elvsteff}.
This figure shows more clearly than Fig.~\ref{fig:onamgal} did that
the distribution function of elemental abundances may
differ in shape between ours and Grundahl et al.'s stars.  For example, Grundahl et al.'s
stars appear to have relatively more O-poor and Al-poor stars than
our sample.  This may be a selection effect since all stars near the RGB tip 
were used whereas the Grundahl et al.\ stars were chosen using the Str\"omgren
$c_1$ index (which Grundahl et al.\ suspected was tracking the N abundance)
to span the extremes in the abundance distribution.
A thorough analysis of a large sample of stars all along the
red giant branch is needed to test this tantalising hint.

This result certainly does not exclude the evolutionary scenario entirely.
Suppose deep mixing occurs over a narrow luminosity range but only in selected
stars.  The result would be the appearance of increased star-to-star abundance
variations at the critical luminosity.  Again, anticipating the discussion of
H-burning products, the mean O and Mg abundance for the more luminous stars should
be lower as the Na and Al abundances should be higher than for stars less luminous
than the critical luminosity.  Such shifts of mean abundances are not evident
within our sample or between our and Grundahl et al.'s samples.  Of course, the
samples are small and there remains the possibility of systematic errors in the
abundance errors masking \teff-dependent effects.  An absence
of a luminosity dependence for the abundances and the presence of similar
star-to-star abundance variations in the main sequence stars, subgiants, and
giants strongly suggests that a primordial not an evolutionary scenario
is responsible for the O, Na, Mg, and Al variations.

Clues to plausible scenarios to account for the star-to-star abundance
variations are provided by consideration of
the extremes of the star-to-star abundance variations.  Knowing that
field (normal) stars fall at the O-rich, Na-poor, Mg-rich, and Al-poor
end of the correlations and anticipating that proton capture reactions
can render gas O-poor, Na-rich, and possibly also Mg-poor and Al-rich,
we dub the O-rich stars `normal' stars.  In the case of the red giants,
the normal stars have the composition of the gas from which they
formed except for rearrangements of C and N introduced by the CN-cycle
and shallow convective mixing.  According to Fig.~\ref{fig:onamgal}, 
the normal stars formed from gas with a composition given by
(O, Na, Mg, Al) $\simeq$ (7.7, 4.6, 6.45, 4.9) or equivalently
([O/Fe], [Na/Fe], [Mg/Fe], [Al/Fe]) $\simeq$ (0.6, $-$0.1, 0.5, 0.0).
These abundances and those of heavier elements 
match well those of a field star of a similar
metallicity.  Extension of this exercise to the Mg isotopes
using Fig.~\ref{fig:isovsal} shows that the normal stars have
abundances of about 6.35: 5.3: 5.3 for $\log\epsilon$($^i$Mg) 
for $i$ = 24, 25, and 26, respectively, corresponding to the relative abundances
$^{24}$Mg:$^{25}$Mg:$^{26}$Mg = 0.8:0.1:0.1.

At the opposite end of the abundance variations to the normal
stars are the O-poor stars which we refer to as `polluted'
stars.  The pollution may originate within a star (the evolutionary
scenario) or have been imprinted on the star 
when it was polluted by ejecta from other
stars (the primordial scenario). 
The compositions of stars at the O-poor end of Fig.~\ref{fig:onamgal} serve to
constrain the composition of the pollutants.  A key factor is 
that the metallicity (iron abundance etc.) is constant across the
sample from normal to polluted stars:
when observational uncertainties are considered, the iron
abundance is independent of (say) the O abundance to 
10 per cent ($\pm$ 0.04 dex) or better.
The invariance of the iron abundance constrains both
evolutionary and primordial scenarios purporting 
to account for the polluted stars. 

In particular, consider a primordial scenario in which
pollutants were contributed by metal-free stars 
from the generation preceding the present stars. 
The observations that the cluster's stars have the same
metallicity to within about 10\% implies that metal-free
pollutants cannot make up more than 10\% of the mass of any
star.  Then, the maximum
underabundance for any element in a polluted star cannot have been
greater than the same 10 per cent.  This is contrary to the observation
that oxygen is underabundant by a factor of 10 in polluted stars. 
Clearly, the pollutants were not metal-free.

Alternatively, suppose that the pollutants were
ejecta from more massive stars of the generation that gave the
present stars.  In this case, the iron abundance of a polluted
star was unaltered by pollution.
In the limit that the pollutants contained no oxygen,
the factor of 10 decline in the oxygen abundance of the
most polluted stars relative to the normal stars
implies that the polluted atmospheres are a mix of
9 parts pollutants to one part normal material. 
The mix shifts in favour of the pollutants in the likely
case that the pollutants contain some oxygen; if the oxygen 
abundance is 10\% of that of normal stars, pollutants
dominate the atmosphere of a polluted star.  For pollutants
with an oxygen abundance greater than 10\% (but less than
100\%), the oxygen abundance of the star falls between that of a
normal and the most polluted star.

This discussion shows that, if pollutants and normal gas have the same
metallicity, compositions of the most O-poor stars are
dominated by pollutants.  It then follows that 
the abundances of elements which are markedly  higher in polluted
than in normal stars reflect the abundances of those elements
in the pollutants.  Sodium, $^{26}$Mg, and Al are in this category.
Figs.~\ref{fig:onamgal} and \ref{fig:isovsal} then 
show that the pollutants are a factor of
about 0.8 dex enriched in Na, 1.0 dex in Al, and 0.6 dex in
$^{26}$Mg relative to the normal stars. 

In the following sections, we discuss the nucleosynthetic
origins of the pollutants in evolutionary and primordial
scenarios.


\subsection{Evolutionary scenario}
\label{sec:ev}

The evolutionary scenario, as defined here, supposes that the star-to-star
differences in C to Al abundances result from internal processing and
both shallow and deep mixing within the red giant.  There is certainly
an evolutionary component to the C and N abundance variations
\citep{ss91}.  Although we have argued that the O, Na, Mg, and Al abundance
variations are primarily of a primordial not an evolutionary origin, we comment
on the principal ideas incorporated into the theoretical proposals for deep mixing
in low mass red giants that appear in discussions of evolutionary scenarios.
Deep mixing is synonymous with H-burning at a high temperature.

Hydrogen-burning at high temperatures qualitatively accounts for the
abundance variations: the CN-cycle turns C to N,
the ON-cycles convert O to N, the NeNa-chain
converts $^{22}$Ne (and $^{20}$Ne) to $^{23}$Na, and the MgAl-chain may burn magnesium
to aluminium.  The Coulomb barrier's height increases from C to Mg
and, thus, the temperature required for efficient burning
increases from C to Mg.  While interior layers with
temperatures necessary for
consumption of C may be readily tapped by non-standard models of
red giants, major revisions of these models are required to link
the convective envelope to deeper hotter
layers where Mg may be converted to Al.  In addition, it is in these
layers that protons are rapidly consumed with the consequence
that material dredged to the surface may alter the He/H of the atmosphere.
As investigated by \citet{dantona02}, He enhancements
will greatly affect the HB morphology.
A review of red giants and possible deep mixing
(evolutionary) scenarios can be found in \citet{salaris02}.
We largely restrict comments to the nucleosynthesis. 

Standard models predict a
maximum temperature of about 55$\times 10^6$ K for the H-burning shell for
a star at the tip of the red giant branch \citep{vsb96}. 
The temperature is lower in less evolved giants.
Higher temperatures have been suggested to result from thermal
instabilities but the \citet{messenger02} assessment 
is that the maximum temperature is likely to be close to 
estimates from standard models. 
Several authors have incorporated deep mixing into their red giant
models and followed the change of surface abundances.  The sense of
the changes may be obtained from calculations of H-burning by 
the simultaneous running of the CNO-cycles,
the Ne-Na and the Mg-Al chains in a layer of uniform temperature
and density reported by \citet{langer95}.
Results were given for T = 40$\times 10^6$ K, a temperature
characteristic of the H-burning shell, and several initial
conditions.  Reduction of the O abundance to its equilibrium
abundance for CNO-cycling  occurs before H is
seriously depleted. Sodium enrichment is dependent on the
conversion of $^{22}$Ne to $^{23}$Na. Plausible values for the `unknown'
abundance of $^{22}$Ne  allow the observed Na abundance
to be  achieved before H exhaustion.
The abundance changes in the atmosphere depend on the
fraction of material in the convective envelope exposed to the
high temperatures. If a large fraction is exposed, the
decrease in the O abundance and the increase in the Na abundance 
approach the values shown by stars at the low-O and high-Na
end of the star-to-star variations \citep{langer93,langer95}. 

A problem arises in extending this evolutionary scenario to
the Mg-Al chain.  At T = 40$\times 10^6$ K, $^{24}$Mg is immune to proton
capture, and $^{27}$Al is produced solely from $^{25}$Mg and $^{26}$Mg with
substantial depletions of these Mg isotopes occurring before
H exhaustion. Significant amounts of radioactive $^{26}$Al
(say, $^{26}$Al/$^{27}$Al $\sim 0.5$) are produced, which, after
a million years or so, is present as $^{26}$Mg.  Our observed 
$^{25}$Mg abundances are independent of the Al abundance and not sharply
declining as predicted \citep{denissenkov98}.  
The $^{26}$Mg abundance increases with Al abundance
but the prediction is that it should probably decline.
In short, mixing into a H-burning shell at T $\sim 40 \times 10^6$ K
will not account for the observed Mg isotopic abundances.  

By activating proton capture on $^{24}$Mg,
the concentrations of $^{25}$Mg and $^{26}$Mg can
be maintained, and
the potential supply of Al increased.
\citet{langer97} show that in H-burning at 70$\times 10^6$ K the 
abundances of $^{25}$Mg and $^{26}$Mg (relative to $^{24}$Mg)
are approximately preserved as the $^{24}$Mg is
depleted.
Langer et al.'s calculations imply a poor fit to the observed
$^{26}$Mg - Al abundance trend.
\citet{langer97} state that observed extreme abundances for M 13,
a cluster with similar star-to-star variations to NGC 6752, are
achievable if the envelope is a 9 to 1 mix of material exposed severely to
70 $\times 10^6$ K and original material. 

The preceding calculations by Langer and colleagues concerned only the
nucleosynthesis and did not attempt to explore how  mixing
to T $\sim 70 \times 10^6$ K might
occur in a red giant. Among a variety of schemes, we
mention a model of `flash-assisted deep mixing' 
presented by \citet{Fujimoto99} and \citet{aikawa01} in an 
attempt to understand the Mg isotopic and Al abundances
reported by \citet{shetrone96a,shetrone96b} for M 13. In this model, hydrogen is
mixed into the He core to temperatures even exceeding 70 $\times 10^6$ K.
The authors
contend that ``our scenario can be distinguished from others by the
fact that $^{25}$Mg and $^{26}$Mg are enriched at the expense of $^{24}$Mg''.
Detailed predictions are not provided but qualitatively the model
for [Fe/H] = $-$1.6 may match our results.
A potential problem is that the model, like all deep mixing 
models tapping into very hot layers, brings helium
from the core into the envelope and the atmosphere.  This helium   
with that produced by H-burning at high temperatures may change the
He/H ratio of the atmosphere by a factor of 2 to 3. This reduces the
atmospheric opacity and enhances the strength of iron (and other) lines.
An abundance analysis made with the
assumption that all stars have the same He/H ratio would then likely
result in a spread in the Fe/H ratios dependent on (say) the Al abundance,
unless there are compensating effect.  This spread is not seen.

The evolutionary scenario in which abundance changes of the light elements
C to Al are entirely
attributable to internal nucleosynthesis and both shallow and deep mixing in
the observed red giants faces major challenges.  The foremost and
seemingly insuperable challenge is that
the star-to-star variation in light element abundances
is found amongst stars from the main sequence to the tip of the red giant
branch without obvious signs of either a difference in the amplitude of the
variations or an onset of changes on the red giant branch associated
with the occurrence of deep mixing.  Shallow convective mixing affecting conversion of
C to N evidently occurs \citep{ss91}.  (Our sample is
dominated by first ascent giants.  A contribution from AGB stars to 
the evolutionary scenario cannot yet be excluded.)  A second challenge is that
there is as yet no {\it ab initio} requirement for deep mixing
to occur down to layers at the temperatures at which the Mg-Al chain
may run.  Even with mixing induced by various artifices, it is not
obvious that deep mixing can account for the Mg isotopic abundances
reported here. 


\subsection{Primordial scenario}
\label{sec:primordial}

The primordial scenario considers the star-to-star abundance variations
to have been imprinted on the present generation of stars at or after their birth.
Given that the range of the variations does not decrease between the main sequence
and the red giant phases, the imprinting cannot have been confined to a thin
skin of the main sequence star but must have affected the
entire main sequence star or a very large fraction of it.  
While abundances of the light elements C - Al show considerable star-to-star
variations, the abundances of heavier elements, commonly represented
by iron, are highly uniform from star to star with the exception of the
Ba and Eu variations seen in M15 \citep{sneden97}.

Uniformity of the metallicity of stars within a globular cluster has
surprisingly evoked far less theoretical attention as measured
by published papers than the
search for an acceptable deep mixing (evolutionary) scenario.  In broad terms
(see \citealt{cayrel86,brown91,brown95,lm96,parmentier99,nmn00}),
it is envisaged that a metal-free (or very metal-poor) cloud of gas forms 
a first generation of stars.  The massive stars of this generation
explode as supernovae.  Their metal-enriched ejecta mix with the remaining
metal-poor cluster gas to form a gaseous shell around the former
star cluster.  Order of magnitude estimates place the metallicity of the shell
in the range exhibited by the Galactic globular clusters.
Similarly, it is argued that the ejecta will be so well mixed into the
primordial gas that the shell will have a uniform composition.
This argument was likely driven by the knowledge that present cluster stars
have a common metallicity.  It is assumed with or 
without supporting justification that the
intermediate and low mass stars first generation 
stars (i.e., stars not dying as supernovae)
are quickly ejected from the cluster.  Certainly, there is no observational
evidence now for a population of lower metallicity stars.  
The present or second generation of stars is considered to form in the shell.

A successful primordial scenario should not only account for the
star-to-star variations of light element abundances but also
explain the composition of the normal stars.  These stars have
a composition that resembles that of field stars of the
same metallicity.
Numerous attempts to explain the compositions of metal-poor 
stars by modelling the Galaxy's chemical
evolution (e.g., \citealt{alc01}) suggest that
the elemental abundances of NGC 6752's `normal' stars
betray the signature of very metal-poor material contaminated by ejecta 
from very metal-poor massive stars dying as Type II supernovae.  
There is, however, at least one 
exception to this conclusion, an exception revealed by our observations 
for the first time.  
Models of very metal-poor supernovae predict very small yields
of the two neutron-rich Mg isotopes, say 
relative abundances  $^{24}$Mg:$^{25}$Mg:$^{26}$Mg = 0.98:0.01:0.01 
\citep{timmes95}.
This is in sharp contrast to Fig.~\ref{fig:isovsal} where the isotopic abundances
of the normal stars are about
$^{24}$Mg:$^{25}$Mg:$^{26}$Mg = 0.8:0.1:0.1. This disparity between observed
and predicted Mg isotopic ratios suggests that either the predicted
yields from supernovae are in error or there was an additional source, one rich in
$^{25}$Mg and $^{26}$Mg, which contributed along with the supernovae to
the contamination of the very metal-poor gas.

The additional source may be intermediate mass (IM) AGB stars
belonging to the first generation of stars.  These stars, with lifetimes
only slightly longer than those of the lowest mass stars to
explode as supernovae, eject gas at low velocities.  Since these
$Z = 0$ IM-AGBs cannot synthesise iron, their ejecta
must be thoroughly mixed throughout the cluster's gas 
before the present (second generation) stars formed in order that
these stars exhibit a common iron abundance.
Evolution of and nucleosynthesis by metal-free ($Z = 0$) IM-AGB stars
has been examined theoretically by \citet{marigo01},
\citet{chieffi01}, and \citet{siess02}.  Of these
studies, only that by \citeauthor{siess02} has considered surface (i.e., ejecta)
abundances of the Mg isotopes.  Synthesis of $^{25}$Mg and $^{26}$Mg but not
$^{24}$Mg occurs primarily from $\alpha$-captures on $^{22}$Ne with 
the latter produced 
from $3\alpha$-processed $^{12}$C,
proton-captures on the $^{12}$C producing $^{14}$N which after two
$\alpha$ captures becomes $^{22}$Ne.  The key to efficient synthesis
of the Mg isotopes is
that the AGB stars near their terminal luminosity experience
proton-captures at the base of the convective envelope, the so called
hot bottomed convective envelope.  A back-of-the-envelope
calculation shows that through a combination of an unexceptional initial mass function
and the yields predicted by Siess et al., the IM-AGBs can supply
sufficient $^{25}$Mg and $^{26}$Mg to raise the low abundances of
$^{25}$Mg and $^{26}$Mg provided by the supernovae to the much higher
levels observed in the normal stars.  The IM-AGBs may also account for
much of the Na and possibly the Al seen in the normal stars. 
These hot bottomed AGB stars have envelopes rich in $^{14}$N.
Supernovae seem incapable of enriching $Z = 0$ gas to the
observed $^{14}$N abundance of the normal second generation stars
\citep{denissenkov97,alc01}.  The IM-AGB ejecta
may also account for the $^{14}$N abundance.

The origin of the pollutants causing the star-to-star abundance
variations for light elements in the present 
stars cannot be the first-generation IM-AGB stars - see Section \ref{sec:obs}.
Second generation IM-AGBs are eligible sources of the
pollutants.  \citet{cottrell81}, who discovered the Na and Al
abundance variations in globular cluster giants from spectra of
NGC 6752's giants, suggested IM-AGBs as sources and
their He-burning shell as the site for the synthesis of Na and Al:
$^{22}$Ne$(\alpha,n)^{25}$Mg and $^{22}$Ne$(\alpha,\gamma)^{26}$Mg is
followed by neutron captures on $^{22}$Ne and the Mg isotopes to make
Na and Al, respectively.  A traditional $s$-process leading to heavy
elements is inhibited by loss of neutrons to nuclei lighter than iron.
This theoretical conclusion is consistent with the fact that Na-rich and
Al-rich cluster giants are not $s$-process enriched (e.g., see \citealt{M5}). 
Association of Na and Al synthesis with the He-burning shell ran into
difficulties when it was subsequently shown that the Na and Al-rich stars
were depleted in oxygen (e.g., see review by \citealt{suntzeff93}) because $^{16}$O is
also synthesised in the He-burning shell.  This difficulty was alleviated
once it was recognized that metal-poor IM-AGBs develop a hot
bottomed convective envelope capable of running the H-burning ON-cycles and
burning O to N.  Reductions of the envelope's O abundance by factors of
10 to 100 are possible depending on the temperatures at the base of the
envelope and the total exposure of envelope material to high temperatures.
Additional processing of light elements - Ne to Al - including participation
of $^{24}$Mg in the Mg-Al chain may occur at the envelope's base.  
Nucleosynthesis by metal-poor IM-AGBs
have been reported by \citet{forestini97} for $Z = 0.005$,
\citet{ventura01,ventura02} for $Z = 2 \times 10^{-4}$, and Karakas \&
Lattanzio (2003, private communication) for $Z = 0.004$ and $0.008$. 
Karakas \& Lattanzio show that low
metallicity IM-AGBs can develop an envelope (i.e., subsequently comprising
the ejecta) rich in
$^{25}$Mg and/or $^{26}$Mg and Al.  (The $Z$ of NGC 6752's stars is a factor
of 2 to 3 less than the lowest value considered by Karakas \& Lattanzio.)
The impression from Fig.\ref{fig:isovsal} is that the pollutants are
rich in $^{26}$Mg but not in $^{25}$Mg.  Production of $^{26}$Mg without
without $^{25}$Mg is implausible for the Mg-Al chain.  What may be happening is
that the $^{25}$Mg abundance is subject to two effects which balance out:
production of $^{25}$Mg from $^{24}$Mg and loss of $^{25}$Mg to $^{26}$Mg.

It is not sufficient to show that second generation IM-AGBs may
have envelopes with a composition closely resembling that inferred 
above for the pollutants necessary to transform a normal to a polluted
star.  One must show that a low mass cluster star can accrete sufficient
mass of ejecta from the IM-AGBs
to be transformed from a normal to a polluted star, and that
this transformation is a common occurrence for the very large number of
stars in the cluster.  Fortunately, \citeauthor{thoul02}'s (2002) detailed
study of the fate of ejecta from IM-AGBs in a globular cluster
shows that accretion of intracluster gas may be an efficient process.
In the case of NGC 6752, Thoul et al. show that
``more than 60\% of the gas ejected by the AGB stars is accreted by the
cluster
stars'' and ``1 $M_\odot$ stars can accrete an appreciable fraction of
their initial mass.  The envelopes of those stars will reflect the 
composition of the intracluster medium rather than the composition of their
interior even if a physical mechanism is at work to induce mixing with the
deeper layers of the star.''  At this level of contamination, it is to be
expected that the star-to-star abundance variations will be very
similar for dwarfs and giants despite the deep convective envelope of
a giant.\footnote{Mass transfer across a binary system is another way for
a low mass star to acquire pollutants. This is the process by which classical
Barium and CH stars in the field are formed.} 

In summary, the second generation IM-AGBs synthesise and
eject the needed pollutants into the intracluster medium which 
the present low mass stars accreted.  Normal stars are but
slightly contaminated by pollutants.  The polluted stars are severely
contaminated.


\section{Concluding Remarks}
\label{sec:conc}

Star-to-star variations of light element (O to Al) abundances among
globular cluster stars are now well documented.  For NGC 6752, the cluster
where Na and Al abundance variations were discovered by \citet{cottrell81}, 
we have extended investigations of abundances for
cluster red giants to the
isotopic ratios of Mg to provide measurements of the
ratios $^{24}$Mg:$^{25}$Mg:$^{26}$Mg in stars sampling the full
range of the abundance variations.  These isotopic ratios proved to
offer new clues to the chemical evolution of the cluster.
In the case of NGC 6752 (and other clusters), there is now strong
evidence that the star-to-star  variations in the O, Na, Mg, and
Al abundances are present in dwarfs, subgiants as well as the
giants for which the initial discoveries were made.  This
suggests that the abundance variations arise in the course of the
chemical evolution of the cluster.     

According to the canonical picture of a globular cluster, the
first generation of stars formed from primordial gas of very low
metallicity ($Z \simeq 0$).  Supernovae ejecta from this generation
mixed with primordial gas to raise the metallicity of the
cluster gas.
A second generation of stars formed from the cooled enriched
gas.  Low mass stars of this generation comprise the present
cluster members.  All first generation stars are presumed to have
been ejected at an early time.

Into this simple picture must be woven ideas to account for the
star-to-star abundance variations.  Current ideas involve the
IM-AGB stars and their ejecta which can be put into the 
intracluster gas on a relatively short timescale.
Our novel results for the Mg isotopic ratios suggest roles
for both first and second generation IM-AGBs. 

Stars with the highest O abundance and lowest Na and Al abundances
(normal stars in our parlance) are considered to have formed from
the cluster's gas after the primordial gas had been thoroughly
mixed with the supernovae ejecta.  Our $^{25}$Mg and $^{26}$Mg
abundances (relative to $^{24}$Mg) are about an order of magnitude
higher than predicted for ejecta from $Z = 0$ massive stars exploding as
supernovae.  One way to increase the $^{25}$Mg and $^{26}$Mg
abundances is by
contamination of the cluster's gas by ejecta from first generation
$Z = 0$ IM-AGBs followed by thorough mixing of the ejecta occurring before
the onset of star formation that led to the second
generation of stars.  The ejecta are likely to have contributed
not only $^{25}$Mg and $^{26}$Mg but also a considerable fraction of the
Na and Al of the normal stars.
Second generation stars dwell in the intracluster medium into
which the IM-AGBs of this generation eject their envelopes at low
velocity. 
Accretion of this gas by lower mass stars pollutes the stars and
introduces the star-to-star light element abundance variations.
 
This cartoon of an evolutionary scenario demands observational
examination.  In the case of NGC 6752, the abundances of light
elements in AGB stars and, especially, in the main-sequence
stars and subgiants should be determined.  Apparent systematic
abundance differences between the abundances 
(O, Na, Mg, Al, and Fe) among giants and
those of the main sequence stars and subgiants reported by 
\citet{gratton01} would, if real, present any evolutionary or primordial scenario
with a challenge.  Of great value would be measurements of the
Mg isotopic ratios for giants of other globular clusters. 
It will be difficult to push the search to clusters lower in
metallicity than NGC 6752 (see \citealt{shetrone96b}) because the MgH
lines will be very weak.  Exploration of giants in several
clusters of NGC 6752's
or higher metallicity is certainly possible with very large
telescopes.  Not only will it be important to see if other clusters
showing a large spread in oxygen abundances (e.g., M 13) show 
the same dependence of the Mg isotopes on the Al (or Na)
abundances, but it will be interesting to determine the
isotopic abundances in clusters (e.g., M 71) where there is
very little star-to-star variation in the Al abundance.
Measurement of the isotopic O abundances using the CO
infrared bands should be made.  Severe ON-cycling in the hot bottomed
envelope of an IM-AGB may produce copious amounts of $^{17}$O:
$^{17}$O/$^{16}$O $\sim 0.1$ \citep{ventura01,ventura02}.  These
observations hold the promise of unravelling the evolution of
the globular clusters and, perhaps, their origin.

Evolutionary paths may differ from cluster to cluster.  Some
paths may end in partial or complete dissolution of the 
cluster at early or late times.
The clusters' origin and evolution may, as many have speculated, be
related to the origin of some halo field stars.  In closing,
we offer the following two comments on the roles assigned to IM-AGBs
and their possible connection to chemical evolution of clusters
and to field halo stars. 

First, if the early evolution of the cluster is rapid, the second generation
of stars may form before the first generation IM-AGBs contaminate 
the gas with $^{25}$Mg, and $^{26}$Mg (also N, Na, and Al).  In this case,
the present (second generation) stars will have the low $^{25}$Mg and
$^{26}$Mg abundances expected of supernovae ejecta.
Observational pursuit of Mg isotopes in other globular clusters may
test this speculation.  Among metal-poor field stars, it is certainly the
case that the speculation would account for the fact that
some field stars have the very low $^{25}$Mg/$^{24}$Mg and $^{26}$Mg/$^{24}$Mg
ratios expected for supernovae ejecta
but others of a similar metallicity have the higher ratios
found here for normal NGC 6752 stars \citep{gl2000}.  Were these
stars with the higher ratios formed in and shed from
a cluster after the ejecta from the first generation IM-AGBs had
been mixed into the gas?
Before answering `yes', an alternative
possibility needs to be investigated, namely, 
some or even all of the stars may be
mass transfer binaries in which a companion IM-AGB provided the
additional $^{25}$Mg and $^{26}$Mg. 

Second, field stars do not show the star-to-star
abundance variations 
seen in NGC 6752 and some other globular clusters 
\citep{pilachowski96,hanson98}.  This difference
implies that the field stars have not come from
globular clusters like NGC 6752.  Two obvious corollaries to
this conclusion are: (i) the field stars did not form in
a closed environment conducive to accretion of ejecta from IM-AGBs
of the same generation; (ii) clusters like NGC 6572 cannot have
provided many field stars.


\begin{acknowledgements}
We thank the referee, Judy Cohen, for helpful comments.
DY is grateful to Inese Ivans for helpful discussions and insights.
DY thanks Peter H{\" o}flich for assistance in running the
syntheses on a cluster of workstations 
financed by the John W. Cox-Fund of the Department of Astronomy at the
University of Texas.  DLL and DY acknowledge support from the Robert A. Welch
Foundation of Houston, Texas.  FG gratefully acknowledges generous financial 
support from the Carlsberg foundation.
This research has made use of the SIMBAD database,
operated at CDS, Strasbourg, France and
NASA's Astrophysics Data System.
\end{acknowledgements}


\end{document}

%% file: 3481.t1.tex
\begin{table*}
{\scriptsize
\caption{Stellar parameters and elemental abundances for program stars \label{tab:abpar}}
\begin{tabular}{llccccccccccccccc}
\hline
Name1 &
Name2 &
RA &
Dec &
V &
\teff &
log g &
$\xi_t$ &
Macro &
Fe {\scshape i} &
Fe {\scshape ii} &
Fe$^{\mathrm{a}}$ &
O &
Na &
Mg &
Al 
\\
 &
 &
(2000) &
(2000) &
 &
(K) &
(cm s$^{-2}$) &
\multicolumn{2}{c}{(km s$^{-1}$)} &
\multicolumn{7}{c}{log$\epsilon$(Species)} \\
\hline
NGC6752-mg0  &  PD1       & 19:10:58 & $-$59:58:07 & 10.70 & 3928 & 0.26 & 2.7 & 5.00 & 5.85 & 6.00 & 5.88 & 7.22 & 5.36 & 6.42 &  5.91 \\
NGC6752-mg1  &  B1630    & 19:11:11 & $-$59:59:51 & 10.73 & 3900 & 0.24 & 2.7 & 4.75 & 5.87 & 6.05 & 5.90 & 7.53 & 5.09 & 6.42 &  5.65 \\
NGC6752-mg2  &  B3589    & 19:10:32 & $-$59:57:01 & 10.94 & 3894 & 0.33 & 2.5 & 6.00 & 5.87 & 6.09 & 5.91 & 7.63 & 4.91 & 6.44 &  5.60 \\
NGC6752-mg3  &  B1416    & 19:11:17 & $-$60:03:10 & 10.99 & 4050 & 0.50 & 2.2 & 5.50 & 5.89 & 5.94 & 5.90 & 7.54 & 4.93 & 6.46 &  5.60 \\
NGC6752-mg4  &  \nodata   & 19:10:43 & $-$59:59:54 & 11.02 & 4065 & 0.53 & 2.2 & 5.50 & 5.89 & 5.96 & 5.90 & 7.45 & 5.00 & 6.43 &  5.73 \\
NGC6752-mg5  &  PD2       & 19:10:49 & $-$59:59:34 & 11.03 & 4100 & 0.56 & 2.1 & 5.00 & 5.90 & 5.96 & 5.91 & 7.50 & 5.04 & 6.48 &  5.57 \\
NGC6752-mg6  &  B2113    & 19:11:03 & $-$60:01:43 & 11.22 & 4154 & 0.68 & 2.1 & 4.50 & 5.90 & 5.94 & 5.91 & 7.68 & 4.85 & 6.51 &  5.40 \\
NGC6752-mg8$^{\mathrm{b}}$  &  \nodata  & 19:10:38 & $-$60:04:10 & 11.47 & 4250 & 0.80 & 2.0 & 5.00 & 5.82 & 5.84 & 5.82 & 7.39 & 4.97 & 6.38 &  5.57 \\
NGC6752-mg9  &  B3169    & 19:10:40 & $-$59:58:14 & 11.52 & 4288 & 0.91 & 1.9 & 5.00 & 5.87 & 5.86 & 5.87 & 7.51 & 4.96 & 6.44 &  5.60 \\
NGC6752-mg10  &  B2575   & 19:10:54 & $-$59:57:14 & 11.54 & 4264 & 0.90 & 1.8 & 4.75 & 5.86 & 5.90 & 5.87 & 7.48 & 4.96 & 6.44 &  5.61 \\
NGC6752-mg12  &  \nodata  & 19:10:58 & $-$59:57:04 & 11.59 & 4286 & 0.94 & 1.8 & 5.25 & 5.87 & 5.92 & 5.88 & 7.71 & 4.60 & 6.42 &  4.92 \\
NGC6752-mg15  &  B2196   & 19:11:01 & $-$59:57:18 & 11.68 & 4354 & 1.02 & 1.9 & 5.25 & 5.90 & 5.89 & 5.90 & 7.47 & 5.02 & 6.48 &  5.55 \\
NGC6752-mg18  &  B1518   & 19:11:15 & $-$60:00:29 & 11.83 & 4398 & 1.11 & 1.8 & 4.75 & 5.90 & 5.89 & 5.90 & 7.53 & 4.90 & 6.45 &  5.42 \\
NGC6752-mg21  &  B3805   & 19:10:28 & $-$59:59:49 & 11.99 & 4429 & 1.20 & 1.8 & 5.25 & 5.90 & 5.92 & 5.90 & 7.08 & 5.28 & 6.38 &  6.01 \\
NGC6752-mg22  &  B2580   & 19:10:54 & $-$60:02:05 & 11.99 & 4436 & 1.20 & 1.8 & 5.00 & 5.89 & 5.91 & 5.89 & 7.25 & 5.33 & 6.44 &  5.82 \\
NGC6752-mg24  &  B1285   & 19:11:19 & $-$60:00:31 & 12.15 & 4511 & 1.31 & 1.9 & 5.00 & 5.88 & 5.84 & 5.87 & 7.69 & 4.59 & 6.43 &  4.95 \\
NGC6752-mg25  &  B2892   & 19:10:46 & $-$59:56:22 & 12.23 & 4489 & 1.33 & 1.7 & 5.25 & 5.90 & 5.92 & 5.90 & 7.66 & 4.85 & 6.49 &  5.34 \\
NGC6752-702$^{\mathrm{b}}$  &   B702    & 19:11:31 & $-$59:54:33 & 10.83 & 4050 & 0.50 & 2.1 & 5.50 & 5.90 & 5.98 & 5.92 &  \nodata  & 5.17 & 6.28 &  6.05 \\
NGC6752-708$^{\mathrm{b}}$  &   B708    & 19:11:31 & $-$59:54:33 & 10.58 & 4050 & 0.25 & 2.2 &  11.00$^{\mathrm{d}}$  & 5.85 & 5.95 & 5.87 & 7.44 & 4.91 & 6.41 &  5.52 \\
NGC6752-A88$^{\mathrm{b}}$  &   A 88     & 19:10:40 & $-$59:54:33 & 10.52 & 4080 & 0.55 & 2.4 & 5.50 & 5.88 & 5.95 & 5.90 & 7.46 & 5.01 & 6.38 &  5.95 \\
HD 103036$^{\mathrm{c}}$  &     &  &  & 8.18 & 4200 & 0.10 & 3.1 & 7.25 & 5.73 & 5.69 & 5.72 & 7.54 & 4.71 & 6.33 &  5.08 \\
HD 141531$^{\mathrm{c}}$  &     &  &  & 9.15 & 4273 & 0.80 & 1.9 & 5.50 & 5.75 & 5.88 & 5.78 & 7.49 & 4.33 & 6.31 &  4.74 \\

\hline
\end{tabular}

}

\begin{list}{}{}
\item[$^{\mathrm{a}}$]Weighted mean of Fe {\scshape i} and Fe {\scshape ii}
\item[$^{\mathrm{b}}$]Spectroscopically determined stellar parameters
\item[$^{\mathrm{c}}$]Comparison field stars
\item[$^{\mathrm{d}}$]This high value is due to the lower resolving power
\end{list}

Note. --- PD1 and PD2 are from \citet{penny86}, the B xxxx names are from \citet{buonanno86}, 
and A 88 is from \citet{alcaino72}.

\end{table*}

%% file: 3481.t2.tex
\begin{table*}
\caption{Abundance dependences on model parameters \label{tab:aberr}}
\begin{tabular}{lccccccc}
\hline
Abundance &
$\Delta$\teff~$\pm$ 50K &
$\Delta$log g $\pm$ 0.2 &
$\Delta$$\xi_t$ $\pm$ 0.2 &
 &
$\Delta$\teff~$\pm$ 50K &
$\Delta$log g $\pm$ 0.2 &
$\Delta$$\xi_t$ $\pm$ 0.2
\\
 &
\multicolumn{3}{c}{NGC6752-mg0$^{\mathrm{a}}$} &
 &
\multicolumn{3}{c}{NGC6752-mg24$^{\mathrm{b}}$} \\
\hline
log$\epsilon$(Fe\,{\scshape i})  & $\pm 0.02$ & $\mp 0.01$ & $\mp 0.02$ &    & $\pm 0.06$ & $\mp 0.01$ & $\mp 0.02$ \\
log$\epsilon$(Fe\,{\scshape ii}) & $\mp 0.09$ & $\pm 0.04$ & $\mp 0.03$ &    & $\mp 0.01$ & $\pm 0.07$ & $\mp 0.03$ \\
log$\epsilon$(Fe)               & $\pm 0.00$ & $\mp 0.01$ & $\mp 0.03$ &    & $\pm 0.04$ & $\pm 0.01$ & $\mp 0.02$ \\
log$\epsilon$(O)                & $\mp 0.01$ & $\pm 0.04$ & $\mp 0.01$ &    & $\pm 0.02$ & $\pm 0.08$ & $\pm 0.00$ \\
log$\epsilon$(Na)               & $\pm 0.05$ & $\mp 0.03$ & $\mp 0.02$ &    & $\pm 0.05$ & $\mp 0.01$ & $\pm 0.00$ \\
log$\epsilon$(Mg)               & $\pm 0.03$ & $\mp 0.03$ & $\mp 0.06$ &    & $\pm 0.06$ & $\mp 0.02$ & $\mp 0.04$ \\
log$\epsilon$(Al)               & $\pm 0.05$ & $\mp 0.02$ & $\mp 0.02$ &    & $\pm 0.04$ & $\mp 0.01$ & $\mp 0.01$ \\
\hline
\end{tabular}

\begin{list}{}{}
\item[$^{\mathrm{a}}$]mg0: \teff=3928 K, log g=0.26 cm s$^{-2}$, $\xi_t$=2.7 km s$^{-1}$
\item[$^{\mathrm{b}}$]mg24: \teff=4511 K, log g=1.31 cm s$^{-2}$, $\xi_t$=1.9 km s$^{-1}$
\end{list}

\end{table*}

%% file: 3481.t3.tex
\begin{table*} 
\caption{Atomic line list 
\label{tab:line}}
\begin{tabular}{lccrclccr}
\hline
Species &
Wavelength (\AA) &
EP (eV) &
log $gf$ &
 &
Species &
Wavelength (\AA) &
EP (eV) &
log $gf$ \\
\hline
O {\scshape i} & 6300.30 & 0.00 &  $-$9.75 &  & Fe {\scshape i} & 5288.53 & 3.69 &  $-$1.51 \\
O {\scshape i} & 6363.78 & 0.02 &  $-$10.25 &  & Fe {\scshape i} & 5321.11 & 4.43 &  $-$1.19 \\
Na {\scshape i} & 4982.83 & 2.10 &  $-$0.91 &  & Fe {\scshape i} & 5322.04 & 2.28 &  $-$3.03 \\
Na {\scshape i} & 5682.65 & 2.10 &  $-$0.71 &  & Fe {\scshape i} & 5373.71 & 4.47 &  $-$0.71 \\
Na {\scshape i} & 5688.22 & 2.10 &  $-$0.40 &  & Fe {\scshape i} & 5466.40 & 4.37 &  $-$0.57 \\
Na {\scshape i} & 6154.23 & 2.10 &  $-$1.56 &  & Fe {\scshape i} & 5487.75 & 4.32 &  $-$0.65 \\
Na {\scshape i} & 6160.75 & 2.10 &  $-$1.26 &  & Fe {\scshape i} & 5522.45 & 4.21 &  $-$1.40 \\
Mg {\scshape i} & 5528.40 & 4.35 &  $-$0.36 &  & Fe {\scshape i} & 5554.90 & 4.55 &  $-$0.38 \\
Mg {\scshape i} & 5711.09 & 4.35 &  $-$1.73 &  & Fe {\scshape i} & 5560.21 & 4.43 &  $-$1.04 \\
Mg {\scshape i} & 6318.71 & 5.11 &  $-$1.97 &  & Fe {\scshape i} & 5567.39 & 2.61 &  $-$2.80 \\
Mg {\scshape i} & 6319.24 & 5.11 &  $-$2.20 &  & Fe {\scshape i} & 5584.77 & 3.57 &  $-$2.17 \\
Al {\scshape i} & 5557.07 & 3.14 &  $-$1.95 &  & Fe {\scshape i} & 5618.63 & 4.21 &  $-$1.26 \\
Al {\scshape i} & 6696.02 & 3.14 &  $-$1.57 &  & Fe {\scshape i} & 5624.02 & 4.39 &  $-$1.33 \\
Al {\scshape i} & 6698.67 & 3.14 &  $-$1.89 &  & Fe {\scshape i} & 5633.95 & 4.99 &  $-$0.12 \\
Fe {\scshape i} & 4817.78 & 2.22 &  $-$3.53 &  & Fe {\scshape i} & 5635.82 & 4.26 &  $-$1.74 \\
Fe {\scshape i} & 4848.88 & 2.28 &  $-$3.40 &  & Fe {\scshape i} & 5638.26 & 4.22 &  $-$0.72 \\
Fe {\scshape i} & 4877.61 & 3.00 &  $-$3.15 &  & Fe {\scshape i} & 5679.02 & 4.65 &  $-$0.77 \\
Fe {\scshape i} & 4896.44 & 3.88 &  $-$1.90 &  & Fe {\scshape i} & 5705.47 & 4.30 &  $-$1.36 \\
Fe {\scshape i} & 4930.32 & 3.96 &  $-$1.20 &  & Fe {\scshape i} & 5731.76 & 4.26 &  $-$1.15 \\
Fe {\scshape i} & 4969.92 & 4.22 &  $-$0.75 &  & Fe {\scshape ii} & 4817.78 & 2.22 &  $-$3.53 \\
Fe {\scshape i} & 5002.79 & 3.40 &  $-$1.44 &  & Fe {\scshape ii} & 4848.88 & 2.28 &  $-$3.40 \\
Fe {\scshape i} & 5029.62 & 3.42 &  $-$1.90 &  & Fe {\scshape ii} & 4877.61 & 3.00 &  $-$3.15 \\
Fe {\scshape i} & 5090.77 & 4.26 &  $-$0.36 &  & Fe {\scshape ii} & 5143.72 & 2.20 &  $-$3.79 \\
Fe {\scshape i} & 5121.64 & 4.28 &  $-$0.72 &  & Fe {\scshape ii} & 5222.39 & 2.28 &  $-$3.68 \\
Fe {\scshape i} & 5143.72 & 2.20 &  $-$3.79 &  & Fe {\scshape ii} & 5322.04 & 2.28 &  $-$3.03 \\
Fe {\scshape i} & 5222.39 & 2.28 &  $-$3.68 &  & Fe {\scshape ii} & 5567.39 & 2.61 &  $-$2.80 \\
\hline
\\
\end{tabular}
\\
\end{table*}

%% file: 3481.t4.tex
\begin{table*} 
\caption{Magnesium isotopic ratios for program stars
($^{24}$Mg:$^{25}$Mg:$^{26}$Mg) \label{tab:iso}}
\begin{tabular}{lccccccccc}
\hline
Name &
\multicolumn{3}{c}{Initial guesses$^{\mathrm{a}}$} &
 &
\multicolumn{3}{c}{Optimum value$^{\mathrm{b}}$} &
 &
Final ratio$^{\mathrm{c}}$
\\
 &
Region 1 &
Region 2 &
Region 3 &
 &
Region 1 &
Region 2 &
Region 3 &
 &
\\
\hline
NGC6752-mg0  & 60:08:32 & 61:10:29 & 66:08:26 & & 60:07:33 & 62:08:30 & 67:09:24 & & 63:08:30 \\
NGC6752-mg1  & 67:09:24 & 71:09:20 & 76:07:17 & & 67:08:26 & 71:09:20 & 78:07:15 & & 71:08:21 \\
NGC6752-mg2  & 77:05:18 & 79:07:14 & 85:05:10 & & 76:04:20 & 82:06:11 & 90:00:10 & & 83:03:15 \\
NGC6752-mg3  & 75:08:17 & 75:10:15 & 80:08:12 & & 75:08:18 & 76:10:14 & 80:09:12 & & 76:09:15 \\
NGC6752-mg4  & 73:08:19 & 73:09:18 & 80:06:14 & & 72:06:22 & 74:07:19 & 80:06:15 & & 74:06:19 \\
NGC6752-mg5  & 75:10:15 & 75:10:15 & 81:08:11 & & 75:09:17 & 75:10:16 & 82:06:13 & & 76:08:16 \\
NGC6752-mg6  & 82:09:09 & 83:10:07 & 88:08:04 & & 83:08:10 & 84:09:07 & 86:08:05 & & 84:08:08 \\
NGC6752-mg8  & 70:20:10 & 71:17:12 & 76:18:06 & & 68:18:14 & 68:20:12 & 71:21:09 & & 68:19:12 \\
NGC6752-mg9  & 78:14:08 & 76:14:10 & Poor S/N$^{\mathrm{d}}$ & & 72:13:15 & 74:16:09 & 68:24:09 & & 72:17:10 \\
NGC6752-mg10  & 72:15:13 & 70:15:15 & 73:13:14 & & 70:13:17 & 67:10:23 & 73:10:17 & & 69:11:19 \\
NGC6752-mg12  & 88:09:03 & 83:11:06 & 83:14:03 & & 86:08:06 & 82:10:08 & 78:15:07 & & 83:10:07 \\
NGC6752-mg15  & 78:07:15 & 76:13:11 & Poor S/N & & 75:13:11 & 76:04:20 & 83:00:17 & & 77:07:16 \\
NGC6752-mg18  & 81:10:09 & 78:12:10 & Poor S/N & & 76:10:14 & 76:16:08 & 77:10:12 & & 77:12:12 \\
NGC6752-mg21  & 60:00:40 & 60:00:40 & Poor S/N & & 64:01:35 & 68:00:32 & \nodata  & & 67:00:33$^{\mathrm{e}}$ \\
NGC6752-mg22  & 60:00:40 & 60:00:40 & Poor S/N & & 62:12:26 & 68:05:27 & \nodata  & & 65:09:26$^{\mathrm{e}}$ \\
NGC6752-mg24  & 85:10:05 & 85:10:05 & Poor S/N & & 82:08:11 & 85:02:14 & 79:22:00 & & 82:08:10 \\
NGC6752-mg25  & 85:10:05 & 85:10:05 & Poor S/N & & 81:06:13 & 85:00:15 & 86:00:13 & & 84:02:14 \\
NGC6752-702   & \nodata  & \nodata  & \nodata  & & 53:08:39 & 51:03:46 & 54:18:29 & & 53:09:39 \\
NGC6752-708   & \nodata  & \nodata  & \nodata  & & 74:16:10 & 80:08:12 & 86:04:10 & & 78:11:11 \\
NGC6752-A88   & \nodata  & \nodata  & \nodata  & & 71:06:23 & 68:15:17 & 71:19:10 & & 70:11:19 \\
HD 103036 & 80:05:15 & 80:05:15 & 92:03:08 & & 90:00:10 & 94:00:06 & 100:0:0 & & 94:00:06 \\
HD 141531 & 91:04:05 & 94:03:03 & 95:03:02 & & 88:05:07 & 93:00:07 & 97:00:03 & & 91:02:06 \\
\hline
\end{tabular}

\begin{list}{}{}
\item[$^{\mathrm{a}}$]Best fit determined by eye
\item[$^{\mathrm{b}}$]Best fit determined by $\chi^2$ analysis
\item[$^{\mathrm{c}}$]Weighted mean of the ratios derived for regions 1, 2, and 3 weighted
                  by the $\chi^2$ errors.  The formal statistical errors are
                  dwarfed by the systematic uncertainties.  We conservatively estimate 
                  errors in $^{24}$Mg:$^{25}$Mg:$^{26}$Mg=(100-b-c):b:c as b $\pm$ 5 and c $\pm$ 5
\item[$^{\mathrm{d}}$]When the S/N was poor, we did not attempt to derive an isotopic ratio via 
                  the traditional method
\item[$^{\mathrm{e}}$]Region 3 was unusuable due to weakness of the lines and low signal-to-noise
\end{list}

Note. --- Region 1 is the line located at 5134.6 \AA, Region 2 is the line located at 5138.7 \AA, and
Region 3 is located at 5140.2 \AA.  

\end{table*}